\documentclass[fleqn,usenatbib]{mnras}

\usepackage{newtxtext,newtxmath}

\usepackage[T1]{fontenc}
\usepackage{ae,aecompl}

\usepackage{graphicx}	
\usepackage{amsmath}	
\usepackage{amssymb}	
\usepackage{subfig}
\usepackage{longtable}

\newcommand{\sbv}{s$_{BV}$ }
\newcommand{\ejm}{M$_{\rm ej}$}

\newcommand{\rflam}{$\overline{\mathcal{F}}_{r_2}$}
\newcommand{\iflam}{$\overline{\mathcal{F}}_{i_2}$}
\newcommand{\bump}{$t_{r_2}$}
\newcommand{\ibump}{$t_{i_2}$}
\newcommand{\Nif}{$\rm ^{56}Ni$\,} 
\newcommand{\Cof}{$\rm ^{56}Co$\,} 
\newcommand{\Fef}{$\rm ^{56}Fe$\,} 

\title[Characterising the secondary maximum in the r-band for Type Ia Supernovae]{Characterising the secondary maximum in the r-band for Type Ia Supernovae: Diagnostic for the ejecta mass}

\author[Papadogiannakis S. et al.]{
Sem\'eli Papadogiannakis,$^{1}$\thanks{E-mail: semeli@fysik.su.se}
Suhail Dhawan,$^{1}$
Roberta Morosin$^{2}$
and Ariel Goobar$^{1}$
\\
$^{1}$Department of Physics, The Oskar Klein Centre, Stockholm University, Alba Nova University Centre, SE-106 91 Stockholm, Sweden\\
$^{2}$Department of Astronomy, Institute for Solar Physics, Stockholm University, Alba Nova University Centre, SE-106 91 Stockholm, Sweden\\
}
\date{Accepted XXX. Received YYY; in original form ZZZ}

\pubyear{2018}

\begin{document}
\label{firstpage}
\pagerange{\pageref{firstpage}--\pageref{lastpage}}
\maketitle

\begin{abstract}
An increase in the number of studied Type Ia supernovae (SNe~Ia) has demonstrated that this class of explosions has a greater diversity in its observables than was previously assumed. The reasons (e.g. the explosion mechanism, progenitor system) for such a diversity remain unknown. Here, we analyse a sample of $r$-band light curves of SNe~Ia, focusing on their behaviour $\sim$ 2-4 weeks after maximum light, i.e. the second maximum. We characterise the second maximum by its timing (\bump) and the integrated flux (\rflam). We find that \bump\ correlates with  the ``colour-stretch'' parameter \sbv, which can be used as a proxy for $^{56}$Ni mass,  and \rflam\, correlates with the transparency timescale, t$_0$. Using \rflam\, for a sample of 199 SNe from the Palomar Transient Factory and intermediate Palomar Transient Factory, we evaluate a distribution on t$_0$ for a sample of SNe~Ia found in an "untargeted" survey. Comparing this distribution to the predictions of t$_0$ ranges from models we find that the largest overlap in t$_0$ values between models and observations is for the sub-Chandrasekhar double detonation models. We also compare our relations between t$_0$ and \rflam\, with that from 1-D explosion models of \citet{GK18} and confirm that \rflam\, can be used as a diagnostic of the total ejecta mass.
\end{abstract}




\section{Introduction}

Type Ia supernovae (SNe~Ia) exhibit diverse observable properties \citep[see][for a review]{Hillebrandt2013,Maguire2017}. SNe~Ia show a great diversity in their spectroscopic display \citep[e.g.][]{Blondin2012,Folatelli2013} and their peak luminosities differ by a factor of $\sim$ 10 \citep[e.g.]
[]{Suntzeff1996,Suntzeff2003,Stritzinger2006,Li2011,Taubenberger2017}. The amount of \Nif\, \citep{Contardo2000} and the total ejected mass \citep{Scalzo2014} also show a wide dispersion. The dispersion in the \Nif\, masses can explain the width-luminosity relation \citep[WLR;][]{Phillips1993} which is used to correct the SN~Ia peak luminosity and use them as distance indicators in cosmology \citep[see][for a review]{Goobar2011,Leibundgut2018}.

Spectroscopically normal SNe~Ia have optical ($B$-band) light curves showing a rise to maximum of $\sim$ 18\, days and a post-maximum decline to an exponentially declining tail. However, at redder wavelengths ($izYJHK$) filters the light curve morphology is markedly different, showing two maxima \citep{Elias1981,Hamuy1996,Meikle2000}. \citet{Kasen2006} explains the emergence of the second maximum as a result of the ionisation transition in the iron-group elements (IGEs) in the ejecta from doubly to singly ionised, leading to a weakening of \ion{Fe}{III} and \ion{Co}{III} lines and a strengthening of \ion{Fe}{II} and \ion{Co}{II} lines \citep[see also][]{Blondin2015}. The timing of the second maximum ($t_2$) is a function of the optical decline rate \citep{Hamuy1996,Dhawan2015} indicating that objects with more synthesized \Nif\, have a later second maximum. Hence, $t_2$ has been used to estimate the \Nif\, mass \citep{Dhawan2016} and to standardize the SNe for distance measurements in cosmology \citep{Shariff2016}.   

In $r$-band, SNe show a weaker second maximum compared the redder filters, akin to a plateau starting at $\sim$ 15 days after maximum. The aim of this study is to characterise the features of the $r$-band light curve and to quantitatively search for relations with global properties of the SN explosion, e.g. the total radioactive \Nif\, mass and total ejecta mass. This is important also to understand the cause of the $r$-band plateau feature, which has not been studied before. With this paper, we aim to fill the gap in light-curve studies with observations in $r$-band.

The onset of large programs to observe samples of SNe~Ia at low-redshift ($z<0.1$) has a led to a large library of $r$-band light curves. Ongoing and recently concluded campaigns, e.g. the Carnegie Supernova Project (CSP-I)  \citep{Contreras2010,Stritzinger2011}, CfA supernova program \citep{Hicken2009}, Palomar Transient Factory \citep[PTF;]{Rau2009}, Carnegie Supernova Project-II \citep{2018arXiv181008213H,2018arXiv181009252P}, Zwicky Transient Facility (ZTF; Graham et al. in preparation), Foundation Supernova Survey \citep{2018MNRAS.475..193F}, have provided and will provide multi-band light curves of SNe~Ia. Hence, this investigation is very timely to understand the characteristics of $r$-band light curve features and how they connect to physical properties of SN explosions. Moreover, studies with theoretical light curves, cross-matched with spectra, have postulated a causal mechanism for the $r$-band plateau. Therefore, a study of the observational properties of SNe~Ia will be highly complementary to the existing theoretical work.

The structure of this paper is as follows: in Section~\ref{sec-data} we describe the datasets used in this study and in Section~\ref{sec-analysis} we detail the methodology for fitting the $r$-band light curves. We present our results in Section~\ref{sec-results} and finally discuss them and conclude in Section~\ref{sec:discussion} and Section ~\ref{sec:conclusion}. 

\section{Data}
\label{sec-data}
In this investigation, we analyse correlations between well-understood decline parameters, bolometric light curve properties and $r$-band properties at late times (note that in this paper late times refers to epochs later than 10 days after maximum light). We use data from the Carnegie Supernova Project (CSP-I)  \citep{Contreras2010,Stritzinger2011} and CfA supernova program \citep{Hicken2009} for analysing the correlations between the different light curve properties. 

The Palomar Transient factory (PTF) and its successor the intermediate transient factory (iPTF) were large field-of-view transient surveys that discovered and obtained light curves for hundreds of SNe~Ia \citep[see][for details]{Papadogiannakis2019}. Unlike the CSP and CfA, which were exclusively follow-up campaigns, PTF/iPTF also discovered the SNe, giving a better control on the selection effects in finding them, important for characterising properties for a population of SNe~Ia \citep{Papadogiannakis2019}. The drawback of this dataset is that it mostly has photometry in one filter, which is the $R$-band for a large fraction of the SNe. Since we want to characterise the $r$-band second maximum, we only use data for SNe in the phase range +10 to +40 \,days (an example is shown in Figure~\ref{fig:example}). 

A summary of the SNe used in this study along with their derived properties of the secondary maximum is presented in the Appendix \ref{section:tables}.

\section{Analysis}\label{sec-analysis}
In this section, we describe the analysis method for evaluating the parameters in the study.
We use two different methods to probe the plateau or secondary maximum in the $r$-band where we determine the time of the plateau (\bump) and calculate the integrated normalised flux around the dates of the plateau (\rflam). We use the same symbols for both the $r$-band and $R$-band used with the CSP-I, CfA and PTF and iPTF survey data respectively. In this analysis, we explore whether these properties of the $r$-band plateau correlate with the global properties of SNe~Ia, e.g. total radioactive nickel mass, total ejecta mass. 
In previous studies it has been shown that the ordering parameter \sbv\,\citep{Burns2014} correlates with the peak bolometric luminosity \citep{Hoeflich2017}. Arnett's rule states that the instantaneous rate of energy deposition from radioactive decay equals the output flux at maximum \citep{Arnett1982}. Hence, we use the \sbv\, parameter as a proxy for the \Nif\, mass in our analyses. In previous studies, it has been shown that the transparency timescale (t$_0$), i.e. the epoch at which the ejecta optical depth is unity, is directly related to the total ejecta mass \citep{Jeffrey1999,Stritzinger2006}. 

Hence, we test for correlations between the properties of the plateau with the ordering parameter \sbv\, and the transparency timescale. Below, we describe the method for evaluating the properties of the plateau, \bump\, and \rflam, as well as the transparency timescale. The ordering parameter \sbv is calculated using the ``colour model'' in the \texttt{SNooPy} light curve fitting software \citep{Burns2014}.

\subsection{Time of secondary maximum}
To estimate the time of the plateau or secondary bump in $r$-band, \bump, we first run Gaussian processes (GP) with a Mat\'ern $\frac{2}{3}$ kernel to get the most likely function, the latent function $\ell$, that describes the data and its uncertainty. 

Gaussian processes is a non-parametric way of predicting the underlying function behind data and works well with unevenly spaced data, such as the one we have. It is also able to predict an error for each given part of the function which sets it apart from other techniques.

To accommodate the GP priors we normalise the fluxes so that the mean error is 1 and compute \bump\, in flux space. To get a better understanding of the error of the \bump\, estimate we then perturb the data points within their errors 100 times using Monte Carlo simulations and get 100 latent functions.  We then compute the derivative of the sampled functions with respect to time and choose the \bump\, as a point with zero derivative and negative second derivative that lies between day +13 and +40 with respect to maximum light. Using the results of this we can then determine the probability that we have encountered a maximum, a shoulder or neither for each SN. If there is an inflection point rather than a bump we calculate the point of inflection and call it a shoulder. 

\begin{figure}
	\centering
	\includegraphics[width=\hsize]{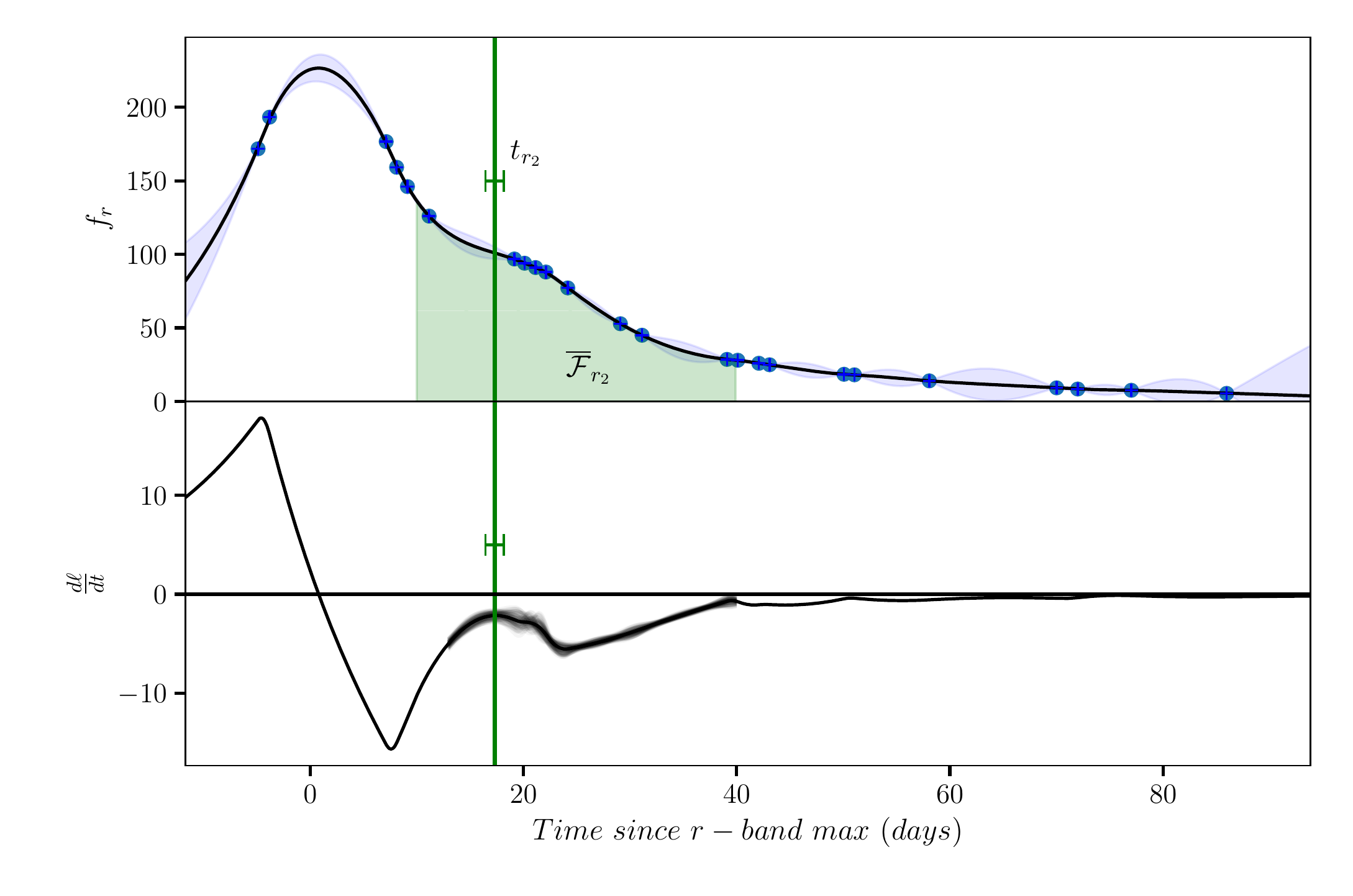}
	\caption{In the upper panel we show an example of the Gaussian processes (GP) fit and estimation of the bump time, \bump\, with the error estimate coming from Monte Carlo (MC) simulations. The green shaded green area shows the estimate of the \rflam. In the lower panel the curve shows the derivative of the most likely function given by GP and the dispersion of the derivative in grey lines. The green solid line shows the estimation of \bump\, averaged over all MC runs.}
	\label{fig:example}
\end{figure}

In Figure~\ref{fig:example} an example fit is shown. We used the same Monte Carlo simulations to estimate the error of \bump.  In our final sample we require the SNe to have at least 4 data points within the times +10 and +40 days with respect to maximum light and a standard deviation of the $t_{max}$ of less than 1.6 days. All SNe rejected by these criteria were visually inspected to make sure no good fit was rejected. This leaves 112 SNe from CfA, 70 SNe from CSP and 240 SNe from PTF and iPTF for which a \bump\, measurement could be obtained. 

\subsection{Integrated flux under the secondary maximum}
\begin{figure}
	\centering
	\includegraphics[width=\hsize]{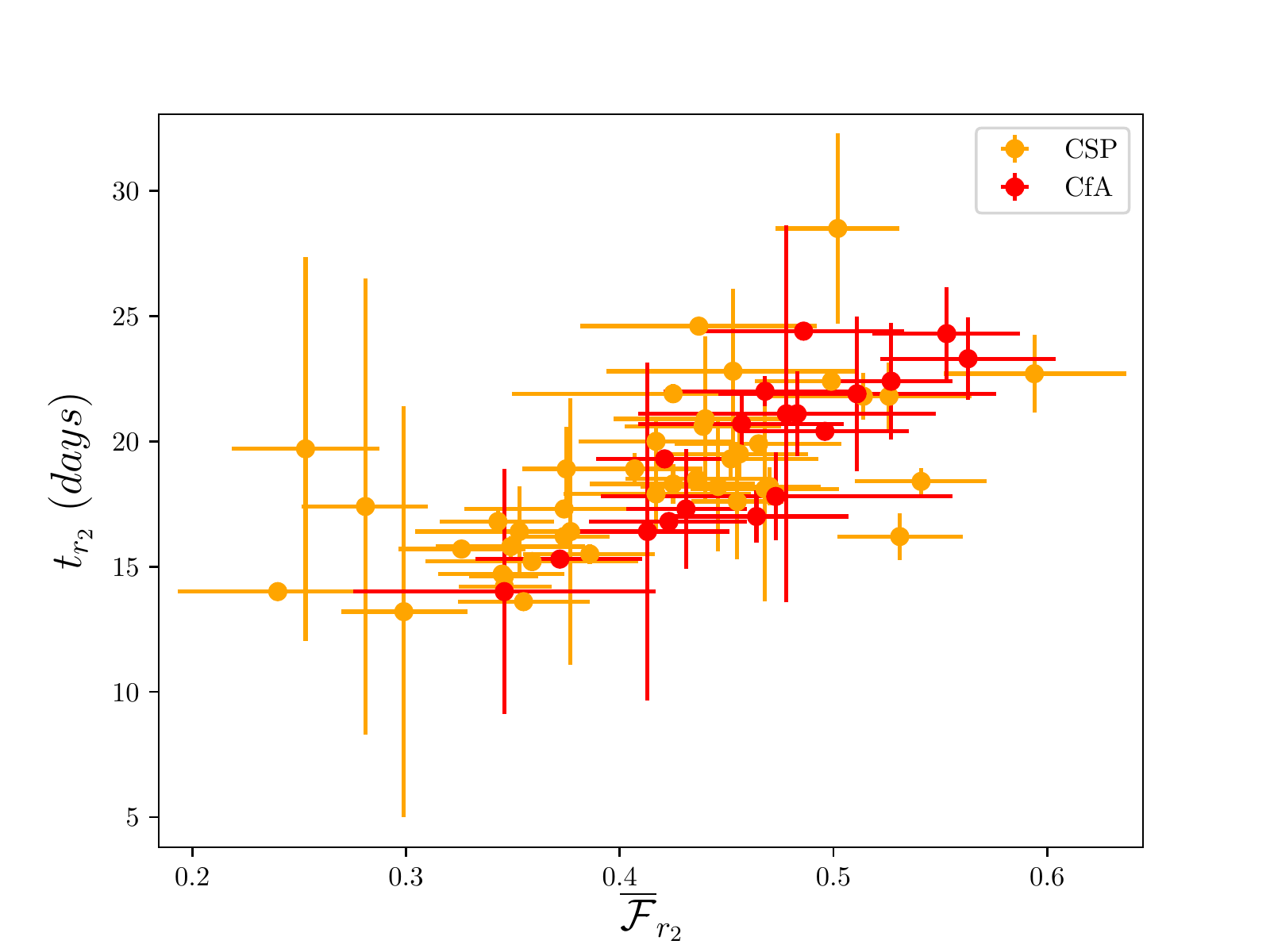}
	\caption{Time of the bump, \bump versus \rflam the two parameters measured at the secondary maximum of the r-band light-curve.}
	\label{fig:bumpflam}
\end{figure}

Another way that we quantify the r-band secondary maximum is using the mean normalised flux integrated in the interval +15 and +40 days with respect to maximum light (\rflam). This metric to quantify the second maximum was proposed by \citet{Krisciunas2001} for the $i$-band, and here, we adapt it for the $r$-band. We calculate \rflam\, by integrating the latent function from heteroscedastic GP of the rest-frame light-curve normalised to the peak flux and divided by the number of days of the interval.  By using GP we get a data driven estimate of the error that allows a robust measurement even when there are gaps in the data. We require at least 4 data points within the integration interval. We get measurements of \rflam\, of 61 SNe from CfA, 53 SNe from CSP and 199 SNe from PTF and iPTF. The two parameters, \rflam\ and \bump are related as seen in figure \ref{fig:bumpflam}.

\begin{figure}
	\centering
	\includegraphics[width=\hsize]{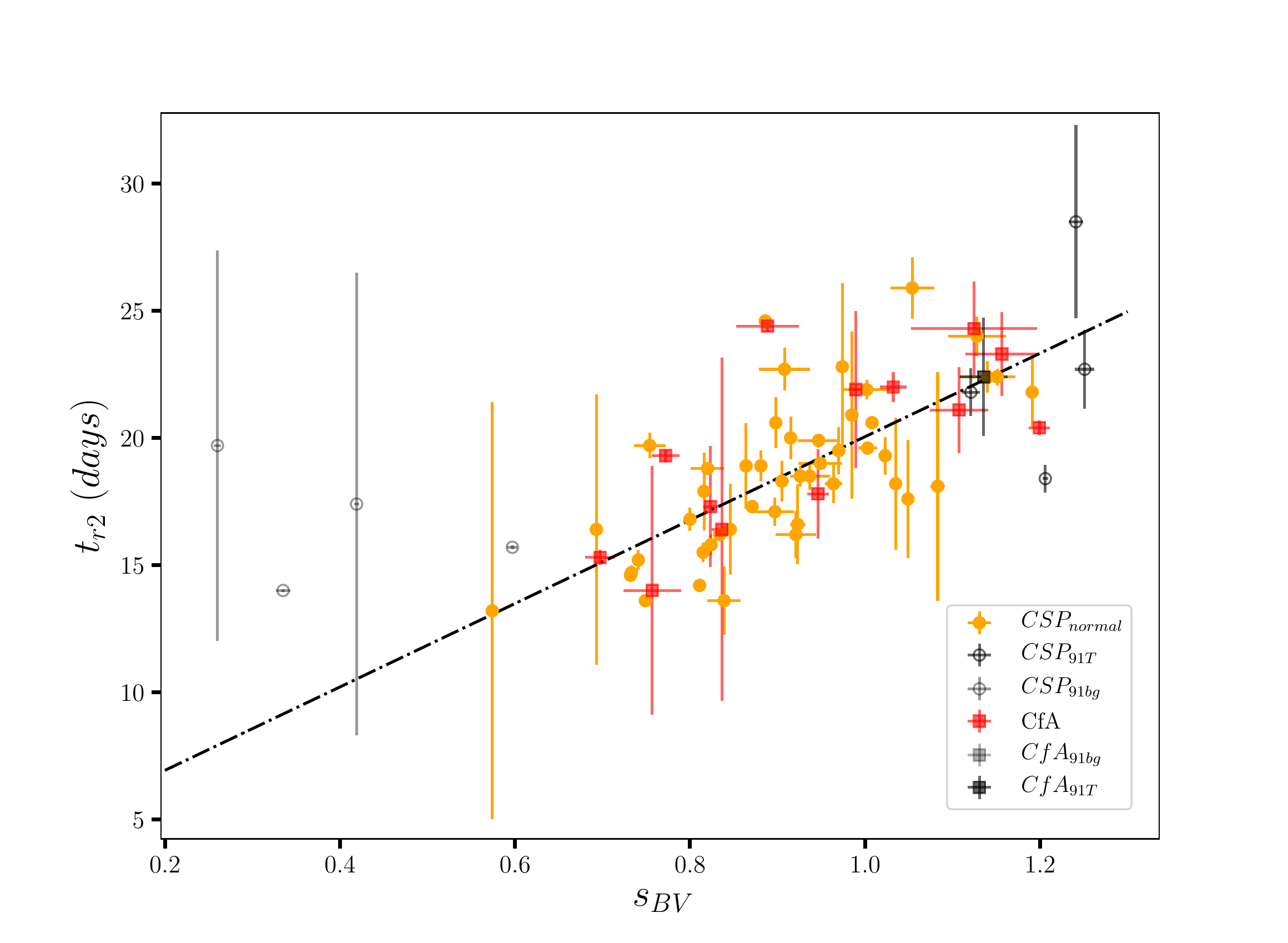}
	\caption{Time of the bump, \bump versus \sbv\, for the SNe from CSP and CfA. The dashed line shows the best fit straight line with a Spearman $r=0.7$ and a $p-value < 10^{-11}$. The gray data points show the spectroscopic outliers such as 91bg and 91T-like SNe from both surveys.}
	\label{fig:sbv_bumptime}
\end{figure}

\begin{figure}
\centering
\includegraphics[width=\hsize]{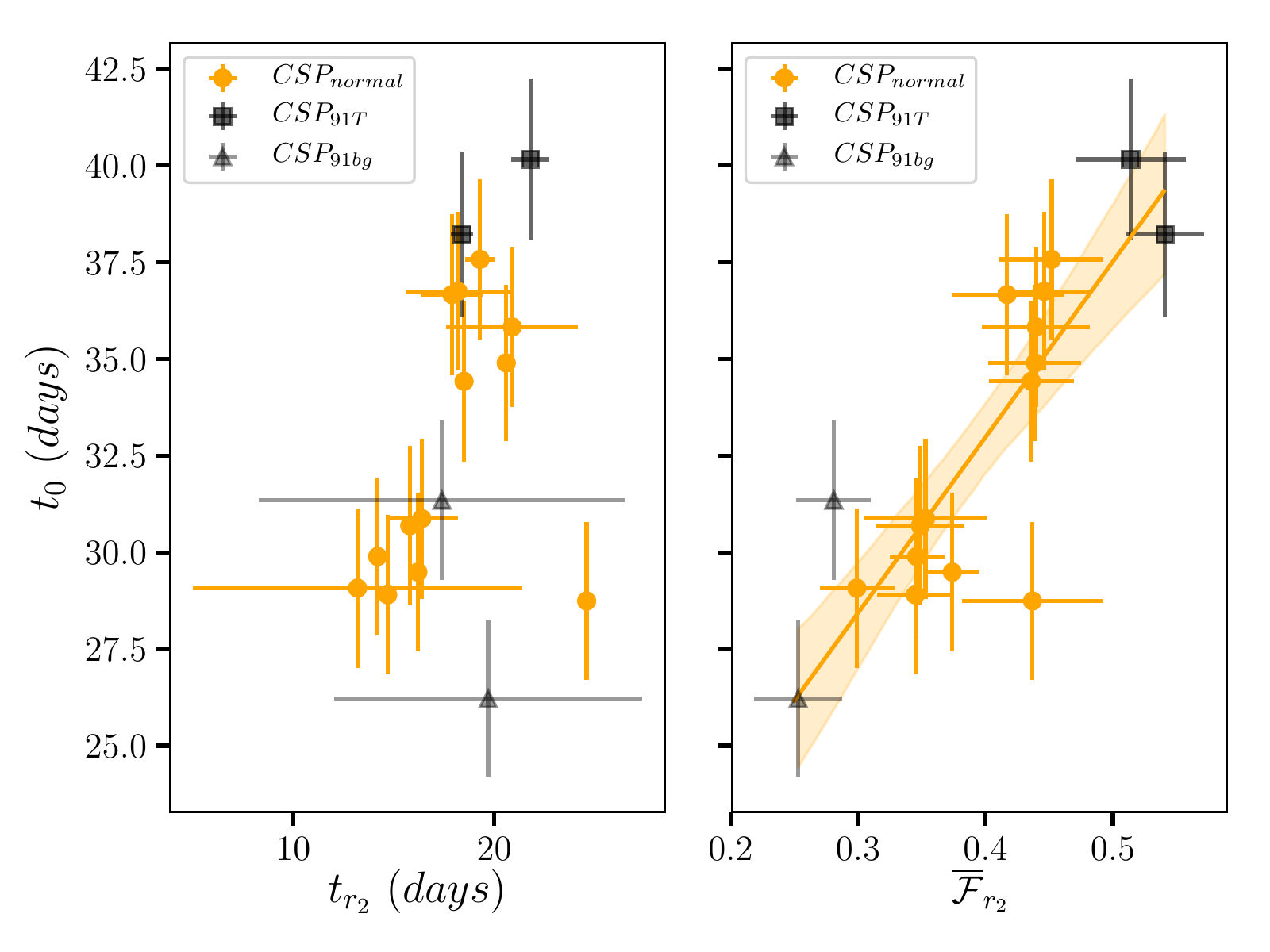}
\caption{\emph{Left}: Transparency timescale t$_0$ vs. bump time, \bump\,  both in units of days. \emph{Right}: Integrated flux, \rflam, versus t$_0$ for the CSP sample, the solid line shows the best fit line and the shaded region the error of the slope. Note that the errorbars in t$_0$ are a systematic error of $2$ days from the uncertainty of the rise-time of Type Ia SNe.}
\label{fig:t0_correlation}
\end{figure}

\subsection{Transparency timescale}
In previous studies \citep{Jeffrey1999,Stritzinger2006}, it has been shown that the transparency timescale, i.e. the epoch at which the ejecta have optical depth of unity, is a proxy for M$_{\rm ej}$. The transparency timescale is evaluated from the bolometric light curve by fitting a radioactive decay energy (RDE) deposition curve to the tail of the observations. We use a standard least squares fitting methodology in this analysis. We add an error of 2\,days to the error from the fit of t$_0$, corresponding to the average error in inferring the rise time of an SN~Ia \citep[see][for details]{Scalzo2014}.
The bolometric light curve is created from the multi-band photometry using the method described in \citet{Dhawan2018}. We convert the observed magnitudes to de-reddened fluxes and interpolate the filters onto the same time step. The fluxes are integrated using the trapezoidal rule and then converted to absolute luminosities using the observed distances \citep[e.g., see][]{Contardo2000}. The expression for the RDE deposition curve is given as follows:

\begin{multline}
\label{eq:dep}
E_{{\rm dep}} =  E_{{\rm Ni}} + E_{{\rm Co~e^{+}}} + [1 - {\rm exp(-\tau_\gamma)}]E_{{\rm Co~\gamma}} \\[2mm]= \lambda_{{\rm Ni}}{{\rm N_{Ni0}}}~{\rm exp(-\lambda_{{\rm Ni}}t)Q_{{\rm Ni~\gamma}}} \\[2mm]+ \lambda_{{\rm Co}}{\rm N_{Ni0}} {\frac{\lambda_{{\rm Ni}}}{\lambda_{{\rm Ni}}-\lambda_{{\rm Co}}}}[[{\rm exp(-\lambda_{{\rm Co}}t)- exp(-\lambda_{{\rm Ni}}t)}]\\[2mm] \times \{Q_{\rm Co~e^{+}} + Q_{{\rm Co~\gamma}}[1 - {\rm exp(- \tau_\gamma)}]\}],
\end{multline}
where the factor (1-exp(-$\tau_\gamma$)) is replaced by 1 for $^{56}$Ni since complete trapping of $\gamma$-rays occurs at early times, when most of the light curve is powered by $^{56}$Ni. $\lambda_{Ni}$ and $\lambda_{Co}$ are the e-folding decay times of 8.8 days and 111.3 days for \Nif\, and \Cof\, respectively.  $Q_{\mathrm{Ni}\,\gamma}$ (1.75 MeV) is the energy release per \Nif\, $\rightarrow$ \Cof\, decay. $Q_{\mathrm{Co}\,\gamma}$ (3.61 MeV) and $Q_{\mathrm{Co}\,e^{+}}$ (0.12 MeV) are the $\gamma$-ray and positron energies, respectively, released per \Cof\, $\rightarrow$ \Fef\, decay \citep[see][]{Stritzinger2006}.  Equation~\ref{eq:dep} is only applicable in the optically thin limit, when the thermalized photons can freely escape.

$\tau_{\gamma}$ is the  mean optical depth, calculated by integrating from the point of emission to the surface of the ejecta \citep[see][for a derivation of the expression]{Jeffrey1999}. It has a simple t$^{-2}$ dependence, given as,
\begin{equation}
\label{eq:tau}
\tau_{\gamma} = \frac{t_0^2}{t^2}.
\end{equation}
 where t$_0$ is the transparency timescale, which by construction in \citet{Jeffrey1999} is the epoch at which the optical depth is unity. 
 
\section{Results}\label{sec-results}

The timing of the NIR second peak has been shown to correlate with light curve properties relating to the peak absolute brightness (e.g. decline rate, $\Delta m_{15}$) \citep{Hamuy1996,Folatelli2010,Biscardi2012,Dhawan2015}. Here, we investigate whether the $r$-band bump shows any significant trends with similar light curve parameters. \citet{Burns2014} demonstrated that the $\Delta m_{15}$ light curve shape parameter does not adequately capture the diversity of SNe~Ia, especially at the faint end, where the SNe transition to the exponential decline at $<$ 15 days from maximum light. As an alternative, they propose a ``colour-stretch'' parameter s$_{BV}$, which is more accurate at ordering even the faint end of the observed distribution of SNe~Ia properties. Previous studies have shown that s$_{BV}$ is correlated strongly with bolometric properties e.g. L$_{\rm max}$ \citep{Dhawan2017a,Hoeflich2017} and hence, a strong indicator of global properties like $^{56}$Ni \citep{Arnett1982}. Thus, we use s$_{BV}$ as a proxy for the intrinsic luminosity of the SN.  

In Figure~\ref{fig:sbv_bumptime}, we plot the resulting correlation between \bump\, and s$_{BV}$ and find a strong correlation between the two quantities (Spearman $r=0.7$ and a $p$-value $< 10^{-11}$). This indicates that SNe with a later \bump\, are intrinsically brighter, similar to the behaviour for the equivalent feature at redder wavelengths \citep{Hamuy1996,Folatelli2010,Dhawan2015}. Hence, \bump\, can be used as a possible luminosity indicator. However when we look at how \bump\, correlates with the Hubble residuals in the Mould $R$-band from the PTF and iPTF sample we find no correlation suggesting that at least in the $R$-band, \bump\ is not a good predictor for luminosity. 

The above correlation relates \bump\ features to the intrinsic luminosity in the B-band. We investigate whether \bump\ also correlates with observables relating to progenitor properties, e.g. the total ejected mass. Previous studies have noted that the 
transparency time-scale (t$_0$) of the bolometric light curve can be an indicator for the ejecta mass \citep{Jeffrey1999,Stritzinger2006,Scalzo2014,Dhawan2017a} with longer time-scales corresponding to higher masses. The transparency time-scale is the epoch when the optical depth of the SN ejecta is unity. We derive it from the bolometric light curve by fitting a radioactive decay energy (RDE) deposition curve to the tail (+40 to +90 days)  of the light curve \citep[see][for details]{Jeffrey1999,Stritzinger2006,Scalzo2014}. 

\begin{figure*}
\includegraphics[width=.8\textwidth]{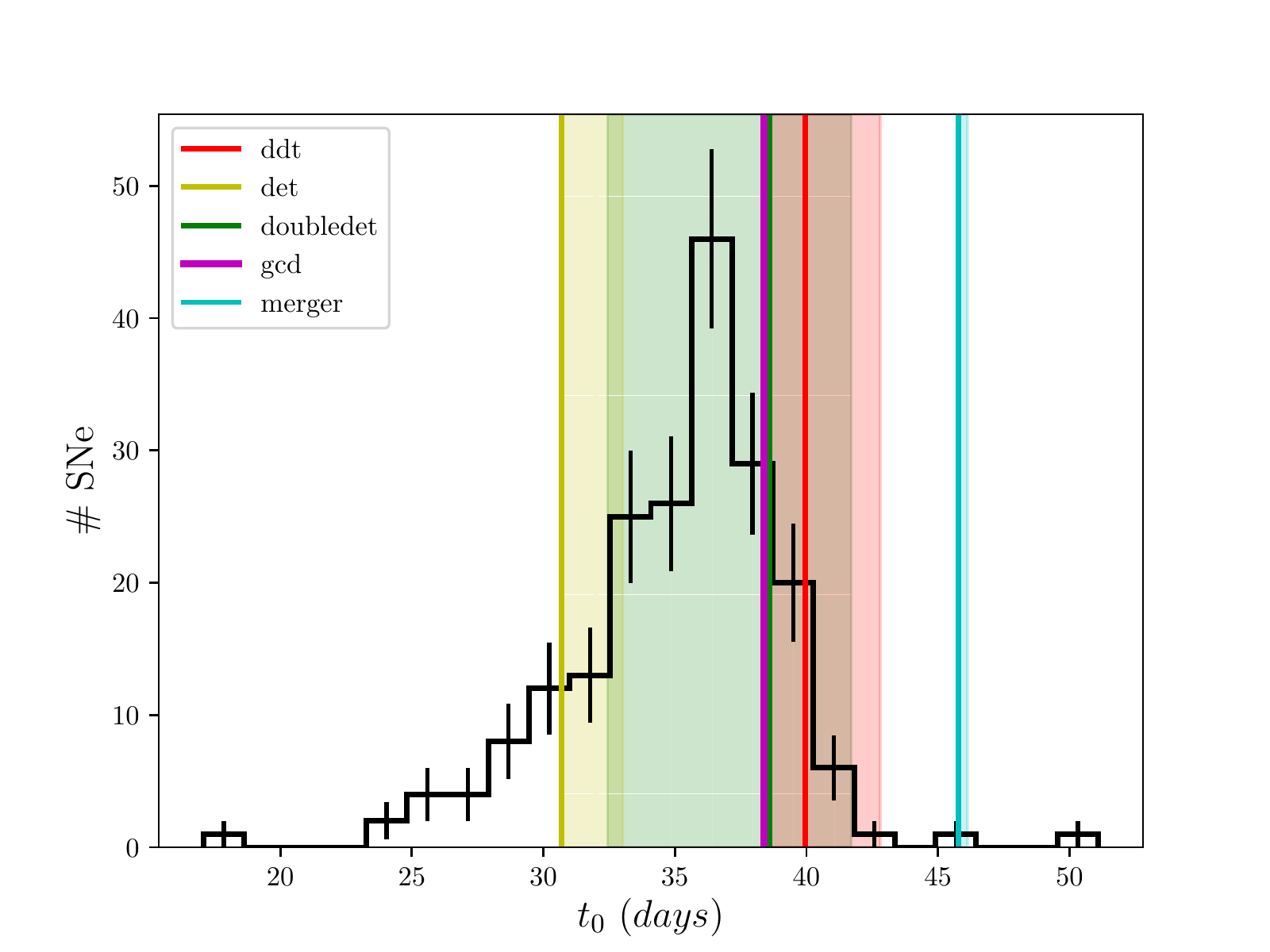}
\caption{This is the extrapolated transparency time-scale t$_0$ distribution of the 199 PTF and iPTF SNe based on the correlation derived from the CSP sample, the errorbars show the Poisson error in each bin. The different shaded regions represent different model prediction of t$_0$.}\label{hist_ptf}
\end{figure*} 

Due to the stringent cuts on the data to sample the peak in $u \rightarrow H$ filters, the final sample of SNe with \rflam\, and t$_0$ measurements is comparatively smaller and coming from the CSP sample only. We note that the small size of the dataset shows the importance of having a parameter, derived only from a single filter, that correlates with t$_0$ to derive t$_0$ values for a large sample of SNe.
We find a significant correlation between the transparency time-scale, t$_0$ and the integrated flux as indicated in Figure~\ref{fig:t0_correlation} with a Spearman $R=0.8$ and a p-value of $<10^{-5}$. The best fit parameters are: 

\begin{equation}
t_0 = 44.92 (\pm 5.86) \times \overline{\mathcal{F}}_{r_2} + 15.00 (\pm 2.32),
\label{eq:t0_fit}
\end{equation}

Since the PTF and iPTF data have a series of well sampled $R$-band light curves (\cite{Papadogiannakis2019}) from an untargeted survey, we can apply Equation~\ref{eq:t0_fit} to a large dataset where the lack of multi-band data would have otherwise prevented us from deriving t$_0$ in the absence of the above relation. We get an extrapolated distribution of t$_0$ shown in Figure \ref{hist_ptf}. The uncertainty on the derived t$_0$ for each object is $\sim$ 3 days, we expect that a larger sample of SNe for deriving the best fit relation will decrease this error.

In previous works a direct comparison of the model predictions for t$_0$ versus other global properties (e.g. \Nif\, mass) has been important to suggest that multiple progenitor channels could be contributing to the observed diversity of SNe~Ia \citep{Scalzo2014,Childress2015,Wygoda2017,Dhawan2018}. Additionally, studies have compared different properties of SNe~Ia (e.g. brightness, \Nif\, mass) with the predictions from models \citep{Ruiter2013,Piro2014}. We present the distribution of t$_0$, such that it can be compared with theoretical predictions for specific model scenarios to distinguish between the different possibilities for the origin of SNe~Ia. In Figure \ref{hist_ptf} we overplot the ranges of t$_0$ from a range of different models taken from the Heidelberg Supernova Model Archive \cite{Kromer17}. The models represented are the pure detonations (det) from \citet{Sim2010}, ``double detonation" (doubledet) from \citet{Fink2010}, gravitationally confined detonation (gcd) from \citet{Seitenzahl2016} and violent merger models (merger) from \citet{Pakmor2010} and \citet{Pakmor2012}. Since the uncertainty on the inferred t$_0$ for the SNe from PTF and iPTF is of order a few days, we do not compare individual SNe to model predictions, but rather compare the range of t$_0$ values observed to the predicted ranges for the different model grids.
 We note that no model covers the entire distribution but the ``double detonation" (doubledet) model is that that covers the largest part of the distribution and the violent merger models cover the least.  In Table~\ref{table:models} we show which percentage of the distribution is covered by each of the tested models.

\begin{table}
\centering 
\begin{tabular}{|l|c|c|}
\multicolumn{2}{c}{}           &  \\ 
\hline
Model     & Fraction &  \\ 
& (\%)\\ \hline
Sub-Chandra Double Detonaton (Doubledet) & 77.4               &  \\
Chandrasekhar mass Delayed Detonation (ddt)       & 17.6               &  \\
Sub-Chandra Detonation (Det)       & 11.6               &  \\
Violent Merger (Merger)    & 0.5                &  \\
\hline
\end{tabular}
\caption{The percentage of SNe in the distribution from \protect\ref{hist_ptf} for the different models tested coming from the Heidelberg Supernova Model Archive \protect\cite{Kromer17}. The gravitationally confined detonations (gcd) scenario is not listed above since it only has one model and hence, we cannot calculate a range of overlap.} \label{table:models}
\end{table}

\begin{figure}
\includegraphics[width=0.5\textwidth]{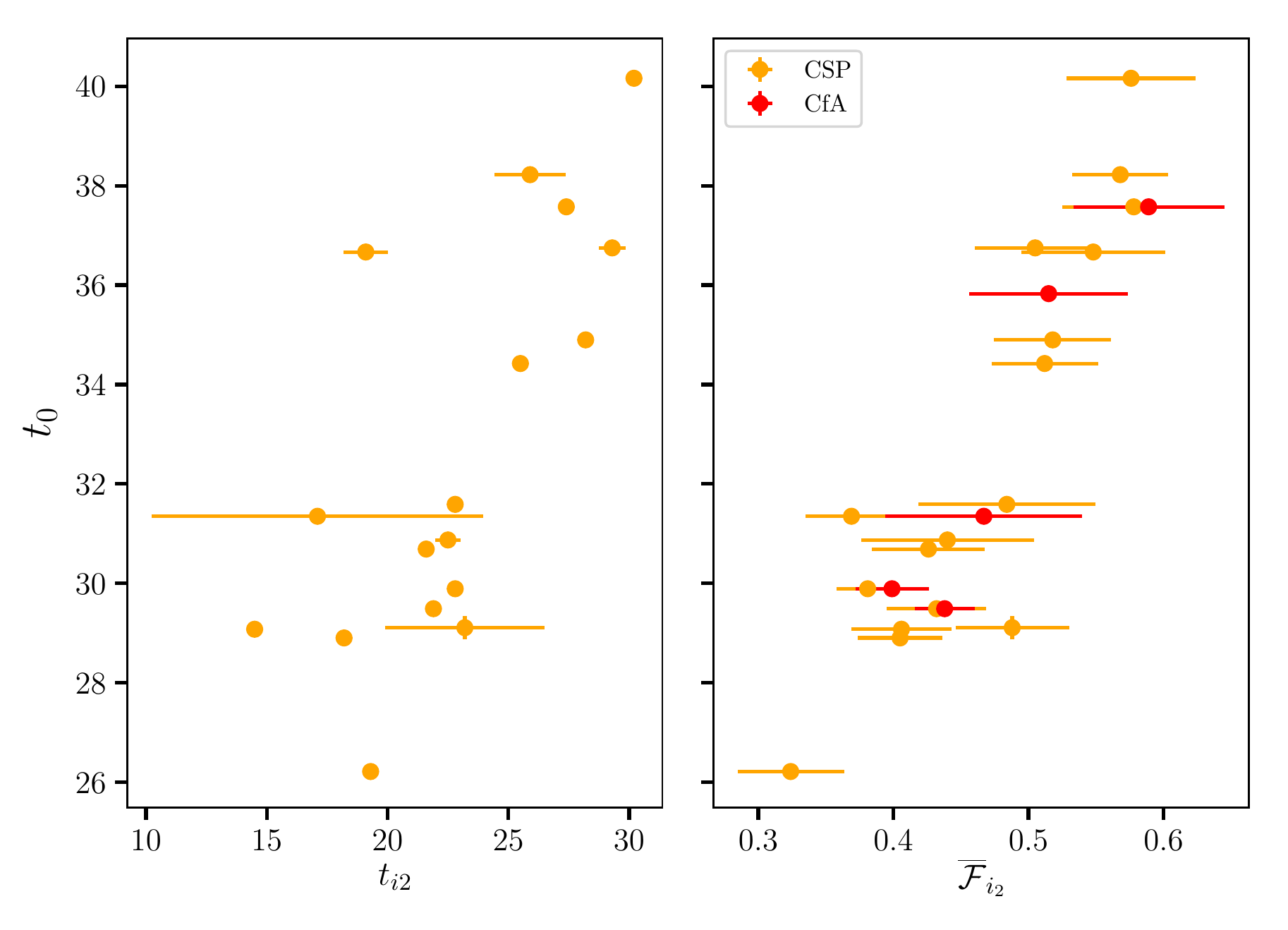}
\caption{I the left panel we show $t_0$ vs. \ibump\ and in the right panel $t_0$ vs. \iflam. We note similar correlations as seen for r-band.}\label{icorr}
\end{figure}

\begin{figure}
\includegraphics[width=0.5\textwidth]{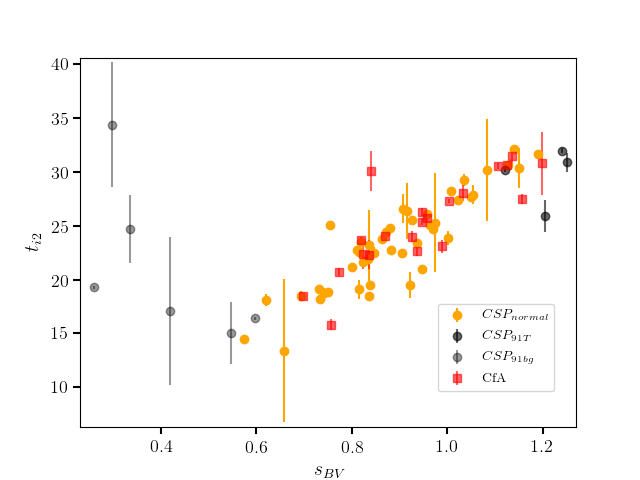}
\caption{Here we show the relation between \sbv and \ibump that is tighter than that in the r-band.}\label{sbvi}
\end{figure}

While our focus here is on the $r$-band, the method can be applied to other filters as well. We apply the same analysis for the $i$-band data of CSP and CfA and find similar correlations as seen in figure \ref{icorr} and \ref{sbvi}.

We note that the dispersion in the \ibump\, vs. \sbv\, relation is tighter in the $i$-band than in the $r$-band whereas the relation between t$_0$ and \iflam\, has similar scatter. This reproduction of correlations seems to indicate a common origin of the $r$- and $i$-band secondary maximum as discussed in \cite{Kasen2006}.

In Figure \ref{rflam_iflam} we show the correlation between the $i$-band and the $r$-band integrated flux, \iflam\, and \rflam.

\begin{figure}
\includegraphics[width=0.5\textwidth]{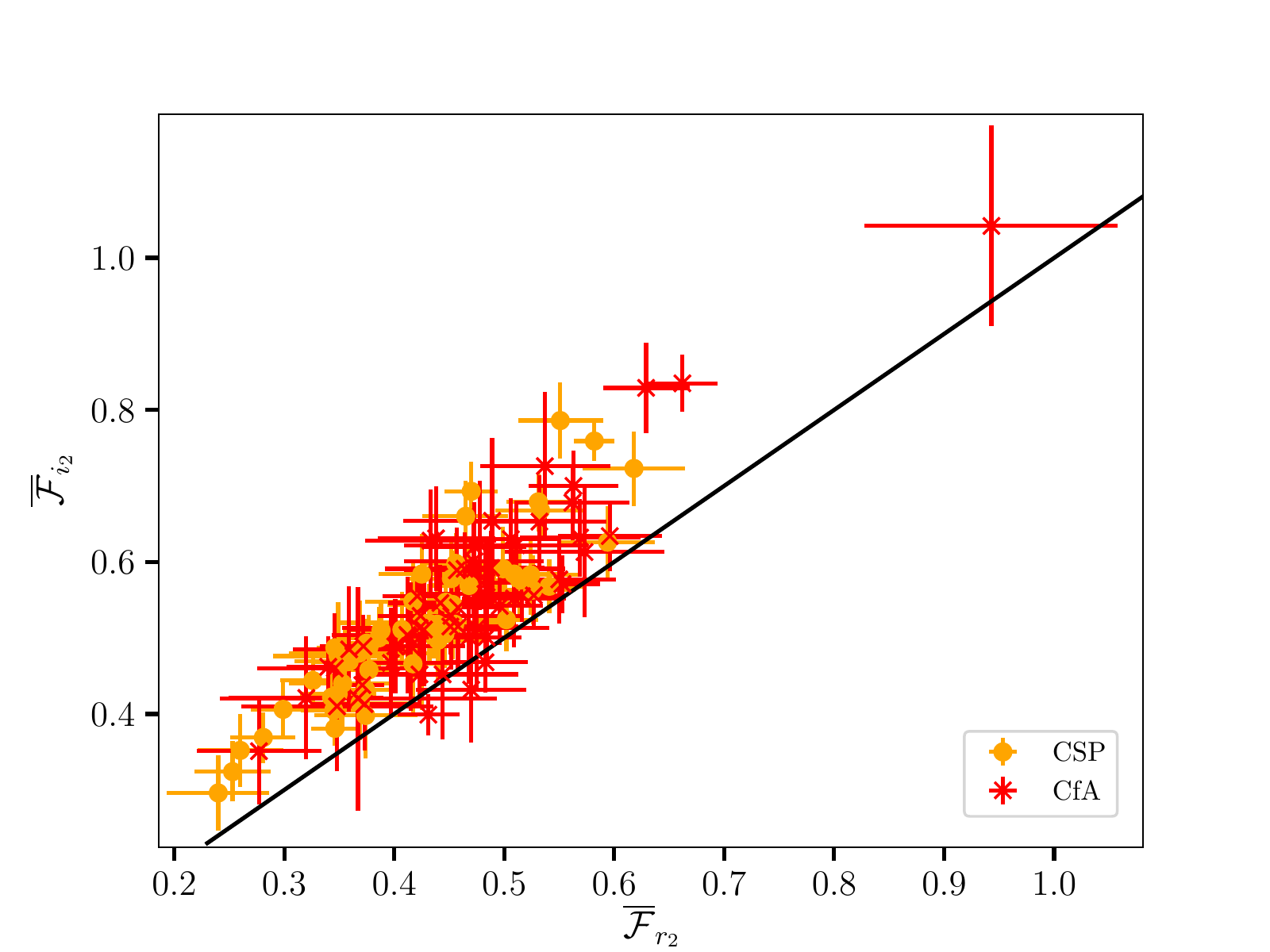}
\caption{\bump\, in days against \rflam\, with the sub-classified SNe in i-band of each survey with different symbols and colours as indicated. The solid black line shows the directly proportional equation between the two parameters.}\label{rflam_iflam}
\end{figure} 

\section{Discussion}\label{sec:discussion}
We find that \bump\ are correlated with s$_{BV}$ as well as the transparency timescale, indicating its link to fundamental properties of the SNe, e.g. luminosity, ejecta mass. We derive equivalent relations for the models and compare them to the observations. 

From the transparency timescale, t$_0$, that we derived earlier, we can get the total ejecta mass (\ejm) using the equation \citep{Jeffrey1999,Stritzinger2006,Dhawan2017a, Dhawan2018};
\begin{equation}
\label{eq:ejm}
M_{\rm ej} = 1.38 \cdot \left(\frac{1/3}{q}\right) \cdot \left( \frac{v_e}{3000\, \mathrm{kms^{-1}}}\right)^2 \cdot \left(\frac{t_0}{36.80\,days}\right)^2 \mathrm{M_{\odot}}.
\end{equation}
for the PTF and iPTF sample. This equation describes the capture rate of $\gamma$ rays in an expanding spherical volume for a given distribution of the radioactive material. Here, $v_e$ is the e-folding velocity, which provides the scaling length for the expansion and $q$ represents the distribution of the $^{56}$Ni in the ejecta. The range of expected v$_e$ values is between $\sim$ 2600 and 3200 km\,s$^{-1}$ with a typical value for a normal SN~Ia being $\sim$ 3000 km\,s$^{-1}$. For the value of $q$, 1/3 implies uniformly distributed ejecta with higher values, till a maximum of unity, implying progressively more centrally concentrated $^{56}$Ni. For the comparison with the relations from the model described below we use $v_e = 3000 km/s$ and $q=1/3$, typical values for a normal SN~Ia (see Figure~\ref{model_rflam}. As in previous studies, we assume a constant $\gamma$-ray opacity of 0.025 ${\rm cm^{2}} {\rm g}^{-1}$ \citep{Swartz1995}.

When comparing with radiative transfer model from \cite{GK18} we find a strong correlation between t$_0$ and \rflam\, suggesting that bump features in the $r$-band can be directly connected to the ejecta mass \ejm.  We analyse a grid of 4500 state-of-the-art radiative transfer models from \cite{GK18} which span a large range of the physical parameters of SNe~Ia, e.g. kinetic energy, total mass, radioactive nickel mass. We compute synthetic photometry in the SDSS-$r$ band and evaluate \bump\, and \rflam\, values for the models. The aim is to note whether the models \emph{directly} show a correlation between M$_{\rm ej}$ (which is M$_{WD}$ for SNe~Ia since the complete white dwarf (WD) is unbound in the explosion) and \bump\, or \rflam, and hence, confirm or refute using theoretical predictions whether the $r$-band bump can be used to derive physical properties for large samples of SNe~Ia. We note that the models of \citet{GK18} are computed under the assumption of local thermodynamic equilibrium (LTE), which does not hold true for the phase of the light curve probed by the bump (see \cite{GK18}), however, they can be used to determine relations between the parameters. 

While we find that the bump like feature in the model is more pronounced than in the data and hence, cannot be used for robust estimates of the theoretical values of \bump\, and \rflam, we can use the inferred values for understanding correlations between the parameters. 
For models with M$_{WD} > 0.5 $ M$_{\odot}$ there is a strong correlation between M$_{WD}$ and \bump. 
Explosion modelling has shown that carbon-oxygen WDs of masses $\gtrsim$ 0.7 M$_{\odot}$ leads to an SN~Ia, the low limit on the mass of the exploding WD. Hence, for comparing observations to models, we look at the mass range (M$_{WD} > $ 0.7 M$_{\odot}$).

We also find that the M$_{WD}$ value is correlated with \rflam (Figure~\ref{model_rflam}). This provides corroborating evidence \rflam\, values can be used as a diagnostic of the \ejm\, for SNe~Ia. This in turn means that we can probe the \ejm\, for higher redshifts compared to a similar analysis in the $i$-band in surveys such as LSST to get an estimate of the \ejm\ for higher redshifts provided a minimal cadence of 5 restframe days in order to get a \bump\, and \rflam\, measurement. However, it is important to note that the range of the two free parameters in equation \ref{eq:ejm} is large and when converting to \ejm\, this gives rise to such large uncertainties that it is not possible to accurately quantify the best fit values. We do see a qualitative trend for both the data and the models from \cite{GK18} in the same direction but not with the same values. 

 \begin{figure}
\includegraphics[width=0.5\textwidth]{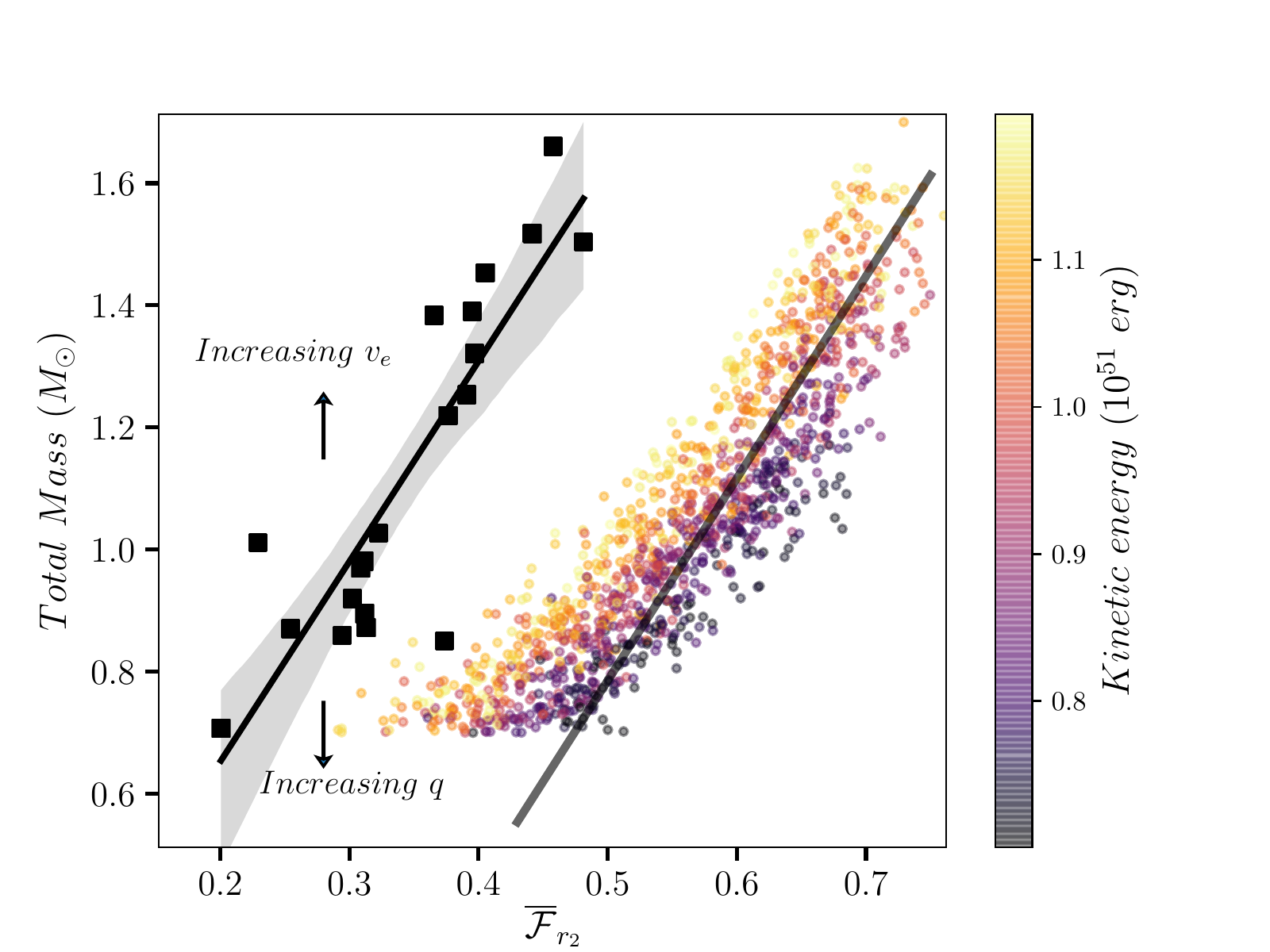}
\caption{The total mass vs. \rflam\, from the \protect\cite{GK18} models. We limit the models shown to the ones with kinetic energy between $0.7-1.2 \times 10^{51}$ erg since that is the physically motivated limits \protect \cite[e.g.][]{Blondin2017}. For comparison we have plotted the total mass vs. \rflam\, for the data (black squares). We derived the total mass from the t$_0$ using typical values of $v_e$ and $q$ from explosion models (see text for more details). The best fit line from the data is shifted using the intercept from the models but keeping the same slope (black line). This shows that the model and the data find similar trends between the total mass and \rflam.}  \label{model_rflam}
\end{figure}

\section{Conclusions}\label{sec:conclusion}
In this paper we present two methods of measuring the secondary maximum of light-curves in the $r$-band that also work for redder bands, e.g. $i$-band. We then used the Carnegie Supernova Project (CSP-I) \citep{Contreras2010,Stritzinger2011} and CfA supernova program \citep{Hicken2009} to derive correlations that we applied to the light-curve sample of PTF and iPTF \citep{Papadogiannakis2019} to show the distribution of the transparency time-scale which is a proxy for total mass. Since the distribution comes from an untargeted survey it represents more closely the true distribution of SNe Ia since the bias of finding SNe in larger galaxies is limited.

In summary our main conclusions are: 
\begin{itemize}
\item We find a significant correlation between \bump\, and \sbv\, which suggest brighter SN have a later bump in the $r$-band, since \sbv\, correlates with the peak brightness of the SN and $^{56}$Ni mass.
\item There is a significant correlation between \rflam\, and the transparency timescale, t$_0$ which is a measure of when the SN ejecta becomes optically thin and is a probe for the total mass of the SN Ia explosion. 
\item These correlations suggest that SN physics can be extracted from the secondary maximum in r-band just as the secondary maximum in the NIR and IR, but with more accessible resources and less observing time.
\item We also find a correlation between \rflam\, and total mass when examining the light-curves in the r-band from the \cite{GK18} models for masses larger than $0.3 \hspace{2mm} M_{solar}$. We note that the shape of the r-band light-curves in the models do not agree with our observations and that the LTE approximations and thus it is only reasonable to compare the trend which is in agreement with our data. More accurate modelling is necessary to further investigate the correlation between \rflam\, and total mass.
\item We see a correlation between the total mass from the modes from \cite{GK18} and \rflam\, strengthening the suggestion that \rflam\, is a probe of the total mass. 
\item When doing the same analysis in the $i$-band similar correlations are found.
\item We present a distribution of the transparency timescale t$_0$ from PTF and iPTF and compare it to models. We find that no model covers the entire range of values from the observations.
\end{itemize}

\section*{Acknowledgements}
We would like to thank D. Goldstein for sending us the model files. We acknowledge support from the Swedish National Science Council as well as the Swedish National Space Agency.








\appendix
\section{Tables}\label{section:tables}
Here we present the tables with the data used to produced the plots in this paper. 
\onecolumn
\begin{table}\caption{The parameters for the CfA SNe in the r-band.}
\begin{tabular}{lcccccccccc}
Name & $t_{r_2}$ & $\sigma_{t_{r_2}}$ & 
$n_{points}$ & sbv & $\sigma_{sbv}$ & Wang & Branch & $\overline{\mathcal{F}}_{r_2}$ & $\sigma_{\overline{\mathcal{F}}_{r_2}}^{lower}$ & $\sigma_{\overline{\mathcal{F}}_{r_2}}^{upper}$ \\
\hline
\\
SN2001ep & 24.4 & 0.21 & 5 & 0.89 & 0.04 & BL & N & 0.486 & 0.44 & 0.53 \\
SN2005am & 15.3 & 0.29 & 11 & 0.70 & 0.02 & BL & N & 0.372 & 0.33 & 0.41 \\
SN2005eq & 22.4 & 2.33 & 15 & 1.14 & 0.03 & SS & 91T & 0.527 & 0.50 & 0.56 \\
SN2005hk & 37.1 & 6.61 & 5 & 0.87 & 0.04 & SS & - & 0.532 & 0.47 & 0.59 \\
SN2006D & 17.3 & 2.38 & 8 & 0.82 & 0.01 & CN & N & 0.431 & 0.40 & 0.46 \\
SN2006N & 19.3 & 0.24 & 10 & 0.77 & 0.02 & CN & N & 0.421 & 0.39 & 0.45 \\
SN2006S & 24.3 & 1.85 & 11 & 1.12 & 0.07 & SS & N & 0.553 & 0.52 & 0.59 \\
SN2006X & 17.0 & 1.04 & 12 & 0.96 & 0.01 & BL & HV & 0.464 & 0.42 & 0.51 \\
SN2006ac & 16.8 & 0.36 & 8 & 0.87 & 0.01 & BL & HV & 0.423 & 0.39 & 0.46 \\
SN2006az & 16.4 & 6.75 & 18 & 0.84 & 0.01 & CN & N & 0.413 & 0.37 & 0.45 \\
SN2006le & 20.4 & 0.3 & 12 & 1.20 & 0.01 & CN & N & 0.496 & 0.46 & 0.54 \\
SN2007co & 17.8 & 1.76 & 11 & 0.95 & 0.01 & BL & N & 0.473 & 0.39 & 0.56 \\
SN2007kk & 23.3 & 1.65 & 13 & 1.16 & 0.04 & BL & N & 0.563 & 0.52 & 0.60 \\
SN2007le & 20.7 & 1.38 & 7 & 1.00 & 0.01 & BL & HV & 0.457 & 0.41 & 0.50 \\
SN2007nq & 14.0 & 4.89 & 6 & 0.76 & 0.03 & BL & N & 0.346 & 0.28 & 0.42 \\
SN2008Z & 21.1 & 1.69 & 8 & 1.11 & 0.03 & SS & N & 0.483 & 0.45 & 0.52 \\
SN2008ae & 21.1 & 7.53 & 9 & 0.84 & 0.07 & SS & - & 0.478 & 0.41 & 0.55 \\
SN2008ar & 21.9 & 3.09 & 5 & 0.99 & 0.01 & CN & N & 0.511 & 0.45 & 0.58 \\
SN2008bf & 22.0 & 0.6 & 9 & 1.03 & 0.02 & CN & N & 0.468 & 0.42 & 0.52 \\
\\
\hline
\end{tabular}

\end{table}
\newpage
\onecolumn
\begin{longtable}{lccccccccccccc}
\caption{The parameters for the CSP-I SNe in the r-band.}\\
Name & $t_{r_2}$ & $\sigma_{t_{r_2}}$ & $n_{points}$ & sbv & $\sigma_{sbv}$ & SNID & Wang & Branch & $z_{Helio}$ & $T_{B(max)}$ & $\overline{\mathcal{F}}_{r_2}$ & $\sigma_{\overline{\mathcal{F}}_{r_2}}^{lower}$ & $\sigma_{\overline{\mathcal{F}}_{r_2}}^{upper}$ \\
\hline
\\
SN2004dt &  &  &  &  &  &  &  &  &  &  & 0.582 & 0.56 & 0.60 \\
SN2004ef & 15.5 & 0.39 & 16 & 0.81 & 0.00 & Normal & HV & BL & 0.03099 & 53264.0 & 0.386 & 0.35 & 0.42 \\
SN2004eo & 17.9 & 1.53 & 16 & 0.82 & 0.01 & Normal & N & CL & 0.0157 & 53278.2 & 0.417 & 0.37 & 0.46 \\
SN2004ey & 20.6 & 0.2 & 11 & 1.01 & 0.00 & Normal & N & CN & 0.01579 & 53303.9 & 0.439 & 0.40 & 0.48 \\
SN2004gc & 16.2 & 0.93 & 11 & 0.92 & 0.02 & Normal & ... & ... & 0.0307 & 53324.6 & 0.531 & 0.50 & 0.56 \\
SN2004gs & 16.4 & 5.32 & 17 & 0.69 & 0.00 & Normal & N & CL & 0.02665 & 53356.0 & 0.377 & 0.35 & 0.40 \\
SN2005A & 18.2 & 0.76 & 18 & 0.96 & 0.01 & Normal & HV & BL & 0.01914 & 53380.4 & 0.47 & 0.45 & 0.49 \\
SN2005M & 18.4 & 0.54 & 17 & 1.21 & 0.00 & 91T & 91T & SS & 0.02462 & 53405.4 & 0.541 & 0.51 & 0.57 \\
SN2005W & 16.6 & 1.57 & 3 & 0.92 & 0.01 & Normal & BL & ... & 0.00889 & 53411.8 &  &  &  \\
SN2005ag & 18.1 & 4.5 & 12 & 1.08 & 0.01 & Normal & N & BL & 0.0798 & 53414.0 & 0.468 & 0.43 & 0.50 \\
SN2005al & 18.9 & 1.69 & 18 & 0.86 & 0.00 & Normal & ... & ... & 0.0124 & 53429.9 & 0.375 & 0.35 & 0.40 \\
SN2005am & 14.6 & 0.27 & 21 & 0.73 & 0.00 & Normal & HV & BL & 0.0079 & 53436.6 & 0.346 & 0.33 & 0.36 \\
SN2005be & 19.7 & 0.5 & 10 & 0.75 & 0.02 & Normal & ... & ... & 0.03502 & 53460.6 &  &  &  \\
SN2005bg & 21.9 & 0.38 & 5 & 1.00 & 0.03 & Normal & N & SS & 0.02309 & 53470.7 & 0.425 & 0.35 & 0.50 \\
SN2005bo & nan & nan & 1 & 0.85 & 0.01 & Normal & N & CN & 0.0139 & 53479.8 &  &  &  \\
SN2005el & 16.2 & 0.27 & 9 & 0.83 & 0.00 & Normal & N & CN & 0.01491 & 53647.1 & 0.374 & 0.35 & 0.40 \\
SN2005eq & 28.5 & 3.8 & 7 & 1.24 & 0.01 & 91T & 91T & SS & 0.02898 & 53654.4 & 0.502 & 0.47 & 0.53 \\
SN2005hc & 21.8 & 1.36 & 7 & 1.19 & 0.01 & Normal & N & CN & 0.04594 & 53667.1 & 0.526 & 0.49 & 0.56 \\
SN2005hj &  &  &  &  &  &  &  &  &  &  & 0.524 & 0.48 & 0.57 \\
SN2005iq & 17.3 & 0.1 & 5 & 0.87 & 0.00 & Normal & ... & ... & 0.03404 & 53687.7 & 0.374 & 0.33 & 0.42 \\
SN2005kc & 20.6 & 1.01 & 3 & 0.90 & 0.01 & Normal & N & CN & 0.01512 & 53697.7 &  &  &  \\
SN2005ke & 17.4 & 9.1 & 9 & 0.42 & 0.00 & 91bg & 91bg & CL & 0.00488 & 53698.3 & 0.281 & 0.25 & 0.31 \\
SN2005ki & 15.8 & 0.41 & 9 & 0.82 & 0.00 & Normal & N & CN & 0.01921 & 53705.6 & 0.349 & 0.31 & 0.38 \\
SN2005ku & nan & 5.51 & 3 & 1.19 & 0.04 & Normal & HV & CN & 0.04544 & 53698.4 &  &  &  \\
SN2005lu & 24.0 & 0.77 & 6 & 1.13 & 0.03 & Normal & ... & ... & 0.03201 & 53711.9 &  &  &  \\
SN2005mc &  &  &  &  &  &  &  &  &  &  & 0.388 & 0.35 & 0.42 \\
SN2005na &  &  &  &  &  &  &  &  &  &  & 0.447 & 0.42 & 0.47 \\
SN2006D & 14.2 & 0.11 & 12 & 0.81 & 0.00 & Normal & N & CN & 0.00853 & 53757.3 & 0.346 & 0.32 & 0.37 \\
SN2006X & 19.5 & 0.93 & 9 & 0.97 & 0.01 & Normal & HV & BL & 0.00524 & 53785.8 & 0.456 & 0.42 & 0.49 \\
SN2006ax & 20.9 & 3.28 & 7 & 0.98 & 0.00 & Normal & N & CN & 0.01674 & 53827.1 & 0.44 & 0.40 & 0.48 \\
SN2006bh & 16.8 & 0.45 & 10 & 0.80 & 0.00 & Normal & ... & ... & 0.01085 & 53833.4 & 0.343 & 0.32 & 0.37 \\
SN2006br & 22.7 & 0.84 & 7 & 0.91 & 0.03 & Normal & HV & BL & 0.02459 & 53851.2 &  &  &  \\
SN2006bt &  &  &  &  &  &  &  &  &  &  & 0.533 & 0.49 & 0.57 \\
SN2006ef & 37.0 & 0.26 & 5 & 0.84 & 0.02 & Normal & HV & BL & 0.01788 & 53969.7 & 0.551 & 0.51 & 0.59 \\
SN2006ej & 18.8 & 0.1 & 5 & 0.82 & 0.02 & Normal & HV & BL & 0.02045 & 53975.6 &  &  &  \\
SN2006eq & nan & 7.06 & 9 & 0.62 & 0.03 & Normal & N & CL & 0.0495 & 53977.1 &  &  &  \\
SN2006et &  &  &  &  &  &  &  &  &  &  & 0.509 & 0.47 & 0.55 \\
SN2006ev & 13.6 & 1.34 & 7 & 0.84 & 0.02 & Normal & ... & ... & 0.02873 & 53989.1 &  &  &  \\
SN2006gj & 35.5 & 4.59 & 5 & 0.66 & 0.01 & Normal & N & CL & 0.02835 & 53999.6 & 0.369 & 0.32 & 0.41 \\
SN2006gt &  &  &  &  &  &  &  &  &  &  & 0.432 & 0.38 & 0.48 \\
SN2006hb & 37.8 & 9.55 & 7 & 0.66 & 0.00 & 86G & 91bg & ... & 0.01534 & 53998.7 &  &  &  \\
SN2006hx & 17.1 & 0.55 & 4 & 0.90 & 0.02 & Normal & N & SS & 0.04549 & 54022.1 &  &  &  \\
SN2006is & 22.4 & 0.62 & 5 & 1.14 & 0.03 & Normal & HV & CN & 0.0314 & 54008.4 &  &  &  \\
SN2006kf & 14.7 & 0.29 & 7 & 0.73 & 0.00 & Normal & N & CL & 0.0213 & 54041.3 & 0.345 & 0.32 & 0.37 \\
SN2006lu & 25.9 & 1.21 & 6 & 1.05 & 0.03 & Normal & ... & ... & 0.0534 & 54034.1 &  &  &  \\
SN2006mr & 19.7 & 7.67 & 9 & 0.26 & 0.00 & 91bg & 91bg & CL & 0.00587 & 54050.1 & 0.253 & 0.22 & 0.29 \\
SN2006ob & 15.2 & 0.39 & 5 & 0.74 & 0.01 & Normal & ... & ... & 0.05924 & 54063.4 & 0.359 & 0.31 & 0.41 \\
SN2006os & 18.5 & 0.54 & 6 & 0.94 & 0.02 & Normal & N & CL & 0.03281 & 54062.8 &  &  &  \\
SN2006ot &  &  &  &  &  &  &  &  &  &  & 0.618 & 0.57 & 0.66 \\
SN2006py & 19.0 & 0.09 & 2 & 0.95 & 0.03 & Normal & ... & ... & 0.0579 & 54070.6 &  &  &  \\
SN2007A & 19.6 & 0.08 & 2 & 1.00 & 0.01 & Normal & N & CN & 0.01765 & 54112.8 &  &  &  \\
SN2007N & 38.4 & 2.91 & 6 & 0.30 & 0.01 & 91bg & 91bg & CL & 0.01288 & 54123.8 & 0.26 & 0.22 & 0.30 \\
SN2007S & 21.8 & 0.94 & 7 & 1.12 & 0.01 & 91T & 91T & SS & 0.01388 & 54144.6 & 0.514 & 0.47 & 0.56 \\
SN2007af & 18.5 & 0.4 & 9 & 0.93 & 0.00 & Normal & N & BL & 0.00546 & 54174.3 & 0.436 & 0.40 & 0.47 \\
SN2007ai & 22.7 & 1.56 & 6 & 1.25 & 0.01 & 91T & 91T & SS & 0.03166 & 54171.8 & 0.594 & 0.55 & 0.64 \\
SN2007as & 18.9 & 0.62 & 6 & 0.88 & 0.00 & Normal & HV & BL & 0.01757 & 54181.3 & 0.407 & 0.38 & 0.44 \\
SN2007ax & 33.8 & 0.41 & 4 & 0.36 & 0.01 & 91bg & 91bg & CL & 0.00686 & 54187.6 &  &  &  \\
SN2007ba &  &  &  &  &  &  &  &  &  &  & 0.368 & 0.30 & 0.43 \\
SN2007bc & 24.6 & 0.12 & 5 & 0.89 & 0.01 & Normal & N & CL & 0.02077 & 54200.1 & 0.437 & 0.38 & 0.49 \\
SN2007bd &  &  &  &  &  &  &  &  &  &  & 0.369 & 0.31 & 0.42 \\
SN2007bm & 18.3 & 0.79 & 5 & 0.91 & 0.01 & Normal & N & CN & 0.00621 & 54224.6 & 0.425 & 0.39 & 0.46 \\
SN2007jg & 20.0 & 0.84 & 8 & 0.92 & 0.01 & Normal & HV & BL & 0.03713 & 54367.2 & 0.417 & 0.38 & 0.45 \\
SN2007jh & 33.4 & 5.59 & 5 & 0.59 & 0.01 & 86G & 91bg & ... & 0.0408 & 54365.3 & 0.349 & 0.29 & 0.41 \\
SN2007le & 19.3 & 0.74 & 6 & 1.02 & 0.00 & Normal & HV & BL & 0.00672 & 54398.7 & 0.452 & 0.41 & 0.49 \\
SN2007nq & 13.6 & 0.12 & 7 & 0.75 & 0.01 & Normal & HV & BL & 0.04503 & 54398.2 & 0.355 & 0.32 & 0.39 \\
SN2007on & 13.2 & 8.2 & 11 & 0.57 & 0.00 & Normal & N & CL & 0.00649 & 54419.7 & 0.299 & 0.27 & 0.33 \\
SN2008C & 19.9 & 0.27 & 6 & 0.95 & 0.02 & Normal & N & SS & 0.01662 & 54466.3 & 0.465 & 0.43 & 0.50 \\
SN2008R & 15.7 & 0.07 & 8 & 0.60 & 0.01 & 91bg & 91bg & CL & 0.0135 & 54494.3 & 0.326 & 0.30 & 0.36 \\
SN2008bc & 18.2 & 2.59 & 11 & 1.03 & 0.00 & Normal & N & CN & 0.01509 & 54548.7 & 0.446 & 0.41 & 0.48 \\
SN2008bq & 22.4 & 0.33 & 6 & 1.15 & 0.01 & Normal & N & CN & 0.034 & 54562.9 & 0.499 & 0.46 & 0.53 \\
SN2008fp & 17.6 & 2.32 & 10 & 1.05 & 0.01 & Normal & N & CN & 0.00566 & 54729.7 & 0.455 & 0.43 & 0.48 \\
SN2008gp & 22.8 & 3.28 & 10 & 0.97 & 0.01 & Normal & ... & ... & 0.03341 & 54778.6 & 0.453 & 0.39 & 0.51 \\
SN2008hv & 16.4 & 1.79 & 11 & 0.85 & 0.00 & Normal & N & CN & 0.01255 & 54816.8 & 0.353 & 0.30 & 0.40 \\
SN2008ia &  &  &  &  &  &  &  &  &  &  & 0.377 & 0.34 & 0.41 \\
SN2009F & 14.0 & 0.05 & 7 & 0.34 & 0.01 & 91bg & 91bg & CL & 0.01296 & 54841.8 & 0.24 & 0.19 & 0.29 \\
SN2009dc &  &  &  &  &  &  &  &  &  &  & 0.666 & 0.61 & 0.72 \\
\\
\hline
\end{longtable}
\newpage

\begin{longtable}{lcccccc}
\caption{The parameters for the CfA SNe in the i-band.}\\
Name & $\overline{\mathcal{F}}_{i_2}$ & $\sigma_{\overline{\mathcal{F}}_{i_2}}^{lower}$ & $\sigma_{\overline{\mathcal{F}}_{i_2}}^{upper}$ &  $t_{i_2}$ & $\sigma_{t_{i_2}}$ & $n_{points}$ \\
\hline
\\
SN2001V & 0.726 & 0.63 & 0.82 & 28.8 & 0.41 & 7 \\
SN2001ep &  &  &  & 18.3 & 1.11 & 4 \\
SN2002bo & 0.654 & 0.55 & 0.76 & 23.3 & 0.37 & 6 \\
SN2002cr &  &  &  & 24.8 & 0.54 & 4 \\
SN2002fk & 0.54 & 0.50 & 0.58 & 26.2 & 0.48 & 6 \\
SN2002ha & 0.489 & 0.40 & 0.57 & 23.7 & 0.29 & 4 \\
SN2002hu &  &  &  & 27.8 & 0.9 & 4 \\
SN2003W & 0.691 & 0.63 & 0.75 & 25.4 & 0.35 & 5 \\
SN2003cg & 0.62 & 0.57 & 0.67 & 24.4 & 0.24 & 5 \\
SN2003du &  &  &  & 28.6 & 0.61 & 4 \\
SN2003kf & 0.515 & 0.48 & 0.55 & 27.4 & 0.61 & 5 \\
SN2005am & 0.489 & 0.45 & 0.53 & 18.5 & 0.12 & 10 \\
SN2005el & 0.438 & 0.42 & 0.46 & 22.2 & 0.15 & 23 \\
SN2005eq & 0.556 & 0.52 & 0.60 & 31.5 & 0.81 & 15 \\
SN2005hc & 0.552 & 0.49 & 0.62 & 30.1 & 1.07 & 10 \\
SN2005hk & 0.653 & 0.59 & 0.71 & nan & 1.38 & 5 \\
SN2005kc & 0.529 & 0.44 & 0.62 & 22.7 & 0.54 & 6 \\
SN2005ke & 0.467 & 0.39 & 0.54 & nan & 8.61 & 7 \\
SN2005ki &  &  &  & 16.2 & 2.24 & 5 \\
SN2005lz & 0.504 & 0.43 & 0.58 & 25.2 & 0.35 & 5 \\
SN2005mc &  &  &  & 31.4 & 8.81 & 4 \\
SN2005ms & 0.501 & 0.46 & 0.55 & 26.4 & 0.59 & 5 \\
SN2005mz & 0.351 & 0.28 & 0.42 & 33.8 & 1.05 & 7 \\
SN2006D & 0.399 & 0.37 & 0.43 & 22.4 & 1.4 & 8 \\
SN2006N & 0.555 & 0.52 & 0.59 & 20.7 & 0.39 & 7 \\
SN2006S & 0.571 & 0.53 & 0.61 & 30.7 & 0.3 & 12 \\
SN2006X & 0.591 & 0.54 & 0.64 & 25.7 & 0.38 & 14 \\
SN2006ac & 0.452 & 0.41 & 0.49 & 24.1 & 0.33 & 9 \\
SN2006ax & 0.515 & 0.46 & 0.57 &  &  &  \\
SN2006az & 0.502 & 0.46 & 0.54 & 22.3 & 1.35 & 18 \\
SN2006bt & 0.678 & 0.63 & 0.73 &  &  &  \\
SN2006cc & 0.511 & 0.44 & 0.59 & 29.5 & 0.33 & 6 \\
SN2006gj & 0.529 & 0.46 & 0.60 & 19.0 & 0.49 & 5 \\
SN2006le & 0.542 & 0.49 & 0.59 & 30.8 & 2.96 & 12 \\
SN2006lf & 0.413 & 0.35 & 0.47 & 20.5 & 0.32 & 8 \\
SN2006ob & 0.41 & 0.32 & 0.49 & 17.3 & 0.29 & 7 \\
SN2006ou &  &  &  & nan & 3.02 & 4 \\
SN2007aj & 0.856 & 0.79 & 0.92 & 24.6 & 0.46 & 10 \\
SN2007bj & 0.829 & 0.77 & 0.89 & 23.8 & 0.27 & 10 \\
SN2007cb &  &  &  & 19.5 & 1.88 & 5 \\
SN2007cf &  &  &  & nan & 3.3 & 6 \\
SN2007cn &  &  &  & nan & nan & 5 \\
SN2007co & 0.591 & 0.50 & 0.68 & 26.3 & 0.34 & 11 \\
SN2007hg & 1.042 & 0.91 & 1.17 & nan & 2.25 & 10 \\
SN2007hj & 0.462 & 0.42 & 0.50 & 15.9 & 0.32 & 9 \\
SN2007hu &  &  &  & 30.1 & 0.89 & 4 \\
SN2007if &  &  &  & nan & 3.31 & 6 \\
SN2007ir & 0.759 & 0.57 & 0.95 & nan & 3.93 & 4 \\
SN2007jg & 0.52 & 0.46 & 0.58 & 27.2 & 0.56 & 9 \\
SN2007kk & 0.7 & 0.65 & 0.75 & 27.5 & 0.49 & 13 \\
SN2007le & 0.589 & 0.53 & 0.65 & 27.3 & 0.19 & 7 \\
SN2007nq & 0.46 & 0.39 & 0.53 & 15.8 & 0.52 & 7 \\
SN2007ob & 0.613 & 0.53 & 0.70 & 28.1 & 0.76 & 6 \\
SN2007ss & 0.546 & 0.50 & 0.60 & 24.1 & 0.43 & 6 \\
SN2007sw & 0.634 & 0.59 & 0.68 & 26.2 & 0.4 & 11 \\
SN2007ux & 0.454 & 0.40 & 0.51 &  &  &  \\
SN2008C & 0.584 & 0.53 & 0.64 & 24.0 & 0.51 & 5 \\
SN2008Y &  &  &  & 28.3 & 1.58 & 6 \\
SN2008Z & 0.468 & 0.43 & 0.51 & 30.6 & 0.27 & 10 \\
SN2008ac &  &  &  & 26.0 & 2.32 & 3 \\
SN2008ae & 0.622 & 0.54 & 0.71 & 30.1 & 1.87 & 8 \\
SN2008ar & 0.622 & 0.56 & 0.68 & 23.1 & 0.6 & 5 \\
SN2008at & 0.707 & 0.63 & 0.78 &  &  &  \\
SN2008bf & 0.501 & 0.44 & 0.56 & 28.1 & 0.82 & 9 \\
SN2008bi &  &  &  & 38.7 & 0.91 & 6 \\
SN2008bw & 0.739 & 0.64 & 0.84 & nan & nan & 5 \\
SN2008cd &  &  &  & 14.2 & 1.95 & 6 \\
SN2008cm & 0.485 & 0.40 & 0.57 & 22.0 & 1.54 & 7 \\
SN2008fr & 0.63 & 0.58 & 0.68 & 29.5 & 0.76 & 10 \\
SN2008gb & 0.546 & 0.47 & 0.62 & 23.8 & 1.28 & 6 \\
SN2008gl & 0.513 & 0.45 & 0.58 & 22.1 & 0.5 & 5 \\
SN2008hm & 0.573 & 0.52 & 0.63 & 24.4 & 0.67 & 10 \\
SN2008hs & 0.628 & 0.56 & 0.70 & 16.4 & 0.26 & 8 \\
SN2008hv &  &  &  & 19.5 & 0.77 & 4 \\
SN2009D & 0.592 & 0.54 & 0.64 & 22.9 & 0.17 & 7 \\
SN2009Y & 0.555 & 0.52 & 0.59 & 28.7 & 0.41 & 12 \\
SN2009ad & 0.558 & 0.50 & 0.62 & 28.0 & 0.32 & 9 \\
SN2009al & 0.632 & 0.58 & 0.68 & 22.0 & 0.45 & 8 \\
SN2009an & 0.489 & 0.43 & 0.55 & 22.2 & 0.39 & 8 \\
SN2009bv & 0.432 & 0.36 & 0.50 & 16.2 & 5.09 & 6 \\
SN2009dc & 0.835 & 0.80 & 0.87 & 20.4 & 1.41 & 14 \\
SN2009do & 0.631 & 0.56 & 0.70 & 23.4 & 0.76 & 8 \\
SN2009ds & 0.452 & 0.37 & 0.54 & 30.8 & 0.46 & 5 \\
SN2009ig & 0.501 & 0.47 & 0.53 & 30.0 & 0.31 & 9 \\
SN2009jr & 0.513 & 0.43 & 0.59 & 31.2 & 0.37 & 8 \\
SN2009kk & 0.529 & 0.49 & 0.57 & 23.5 & 0.49 & 7 \\
SN2009kq & 0.564 & 0.52 & 0.60 & 27.0 & 0.23 & 5 \\
SN2009lf & 0.601 & 0.53 & 0.67 & 23.3 & 0.55 & 6 \\
SN2010Y & 0.421 & 0.34 & 0.50 & 16.6 & 0.31 & 8 \\
SN2010ag & 0.577 & 0.52 & 0.63 & 27.7 & 0.98 & 6 \\
SN2010ai & 0.42 & 0.27 & 0.57 & 22.8 & 3.5 & 6 \\
\\
\hline
\end{longtable}

\newpage
\begin{longtable}{lcccccc}
\caption{The parameters for the CSP-I SNe in the i-band.}\\
Name & $\overline{\mathcal{F}}_{i_2}$ & $\sigma_{\overline{\mathcal{F}}_{i_2}}^{lower}$ & $\sigma_{\overline{\mathcal{F}}_{i_2}}^{upper}$ &  $t_{i_2}$ & $\sigma_{t_{i_2}}$ & $n_{points}$ \\
\hline
\\
SN2004dt & 0.759 & 18.322 & 19.63 &  &  &  \\
SN2004ef & 0.5 & 11.463 & 13.531 & 22.6 & 0.2 & 16 \\
SN2004eo & 0.548 & 12.364 & 15.027 & 19.1 & 0.91 & 16 \\
SN2004ey & 0.518 & 11.854 & 14.032 & 28.2 & 0.14 & 11 \\
SN2004gc & 0.679 & 16.01 & 17.915 & 19.5 & 1.21 & 10 \\
SN2004gs & 0.458 & 10.732 & 12.148 & 18.5 & 0.46 & 16 \\
SN2005A & 0.693 & 16.33 & 18.308 & 25.1 & 0.18 & 18 \\
SN2005M & 0.568 & 13.313 & 15.09 & 25.9 & 1.46 & 17 \\
SN2005ag & 0.569 & 12.974 & 15.485 & 30.2 & 4.72 & 12 \\
SN2005al & 0.493 & 11.507 & 13.141 & 23.8 & 0.14 & 18 \\
SN2005am & 0.487 & 11.474 & 12.888 & 19.1 & 0.14 & 19 \\
SN2005be & 0.737 & 16.616 & 20.245 & 25.1 & 0.25 & 10 \\
SN2005bg & 0.52 & 10.126 & 15.869 & 23.9 & 0.66 & 5 \\
SN2005el & 0.432 & 9.877 & 11.711 & 21.9 & 0.11 & 9 \\
SN2005eq & 0.524 & 12.064 & 14.148 & 32.0 & 0.26 & 7 \\
SN2005hc & 0.563 & 12.944 & 15.224 & 31.7 & 0.21 & 7 \\
SN2005hj & 0.583 & 13.268 & 15.878 & 32.7 & 0.11 & 5 \\
SN2005iq & 0.398 & 8.525 & 11.366 & 24.4 & 0.23 & 5 \\
SN2005ir &  &  &  & 26.5 & 0.3 & 4 \\
SN2005kc &  &  &  & 22.6 & 0.16 & 4 \\
SN2005ke & 0.369 & 8.38 & 10.055 & 17.1 & 6.87 & 9 \\
SN2005ki & 0.426 & 9.605 & 11.688 & 21.6 & 0.14 & 9 \\
SN2005ku &  &  &  & 28.3 & 1.12 & 3 \\
SN2005lu & 0.779 & 18.17 & 20.775 & 30.7 & 0.34 & 6 \\
SN2005mc & 0.51 & 11.792 & 13.7 & 17.8 & 0.23 & 11 \\
SN2005na & 0.547 & 12.67 & 14.682 & 26.1 & 0.37 & 11 \\
SN2006D & 0.381 & 8.949 & 10.078 & 22.8 & 0.2 & 12 \\
SN2006X & 0.598 & 13.929 & 15.947 & 24.7 & 0.13 & 9 \\
SN2006ax & 0.497 & 11.151 & 13.686 &  &  &  \\
SN2006bd &  &  &  & nan & 0.06 & 5 \\
SN2006bh & 0.422 & 9.787 & 11.33 & 21.2 & 0.1 & 10 \\
SN2006br & 0.696 & 15.594 & 19.196 & 26.6 & 1.35 & 7 \\
SN2006bt & 0.668 & 15.846 & 17.543 &  &  &  \\
SN2006ef & 0.786 & 18.396 & 20.916 & 18.5 & 0.15 & 5 \\
SN2006ej & 0.573 & 13.322 & 15.322 & 23.5 & 0.18 & 5 \\
SN2006eq & 0.531 & 12.069 & 14.474 & 18.1 & 0.58 & 9 \\
SN2006et & 0.584 & 13.672 & 15.529 &  &  &  \\
SN2006ev & 0.488 & 10.689 & 13.719 & 19.5 & 0.26 & 7 \\
SN2006gj & 0.414 & 9.132 & 11.583 & 13.4 & 6.65 & 5 \\
SN2006gt & 0.519 & 11.658 & 14.27 &  &  &  \\
SN2006hb & 0.733 & 17.354 & 19.304 & 18.8 & 4.21 & 7 \\
SN2006is & 0.622 & 14.411 & 16.67 & 32.1 & 0.18 & 5 \\
SN2006kf & 0.405 & 9.347 & 10.902 & 18.2 & 0.18 & 7 \\
SN2006lu & 0.657 & 15.086 & 17.751 & 27.9 & 0.91 & 6 \\
SN2006mr & 0.324 & 7.119 & 9.099 & 19.3 & 0.14 & 9 \\
SN2006ob & 0.469 & 10.28 & 13.192 & 18.8 & 0.42 & 5 \\
SN2006os & 0.605 & 13.954 & 16.272 & 23.4 & 0.37 & 6 \\
SN2006ot & 0.723 & 16.846 & 19.293 &  &  &  \\
SN2007N & 0.352 & 7.6 & 10.0 & 34.4 & 5.82 & 6 \\
SN2007S & 0.576 & 13.206 & 15.593 & 30.2 & 0.17 & 7 \\
SN2007af & 0.512 & 11.816 & 13.795 & 25.5 & 0.33 & 9 \\
SN2007ai & 0.626 & 14.482 & 16.833 & 30.9 & 0.85 & 6 \\
SN2007as & 0.511 & 11.552 & 14.016 & 24.8 & 0.13 & 6 \\
SN2007ba & 0.48 & 10.313 & 13.691 & 15.0 & 2.88 & 7 \\
SN2007bc &  &  &  & 23.2 & 3.56 & 5 \\
SN2007bd & 0.484 & 10.464 & 13.735 & 22.8 & 0.17 & 5 \\
SN2007bm & 0.584 & 13.27 & 15.922 & 22.5 & 0.15 & 5 \\
SN2007ca &  &  &  & 29.5 & 0.32 & 4 \\
SN2007if &  &  &  & 22.5 & 0.68 & 6 \\
SN2007jg & 0.466 & 9.929 & 13.378 & 26.4 & 2.62 & 7 \\
SN2007jh & 0.476 & 10.109 & 13.676 & 14.4 & 4.58 & 5 \\
SN2007le & 0.578 & 13.133 & 15.782 & 27.4 & 0.3 & 6 \\
SN2007nq & 0.468 & 10.742 & 12.675 & 18.9 & 0.19 & 7 \\
SN2007on & 0.406 & 9.221 & 11.072 & 14.5 & 0.25 & 11 \\
SN2008C & 0.66 & 15.328 & 17.678 & 21.0 & 0.31 & 6 \\
SN2008R & 0.444 & 10.25 & 11.936 & 16.4 & 0.13 & 8 \\
SN2008bc & 0.505 & 11.511 & 13.751 & 29.3 & 0.56 & 11 \\
SN2008bq & 0.591 & 13.387 & 16.163 & 30.4 & 1.84 & 6 \\
SN2008fp & 0.518 & 12.181 & 13.709 & 27.7 & 0.17 & 10 \\
SN2008gp & 0.544 & 11.801 & 15.378 & 25.3 & 4.59 & 10 \\
SN2008hv & 0.44 & 9.412 & 12.602 & 22.5 & 0.52 & 11 \\
SN2008ia & 0.488 & 11.149 & 13.261 & 23.2 & 3.31 & 8 \\
SN2009F & 0.296 & 6.165 & 8.653 & 24.7 & 3.14 & 7 \\
SN2009dc &  &  &  & 20.8 & 0.72 & 4 \\
\\
\hline
\end{longtable}

\newpage
\onecolumn
\begin{table}\caption{The $t_0$ values from the CSP-I SNe with sufficient multi-band data.}
\begin{tabular}{ccc}
Name & $t_0$& $\sigma_{t_0}$ \\
\hline
\\
SN2004eo & 36.6655 & 0.0802 \\
SN2004ey & 34.8975 & 0.0284 \\
SN2005M & 38.2209 & 0.1411 \\
SN2005el & 29.4895 & 0.0517 \\
SN2005ke & 31.3489 & 0.0548 \\
SN2005ki & 30.6891 & 0.0613 \\
SN2006D & 29.8909 & 0.0486 \\
SN2006ax & 35.8279 & 0.0738 \\
SN2006et & 38.3934 & 0.0918 \\
SN2006kf & 28.9012 & 0.0587 \\
SN2006mr & 26.2137 & 0.0234 \\
SN2007S & 40.1609 & 0.0925 \\
SN2007af & 34.4224 & 0.0766 \\
SN2007ax & 27.8355 & 0.0396 \\
SN2007bc & 28.7434 & 0.046 \\
SN2007bd & 31.5881 & 0.0021 \\
SN2007le & 37.5743 & 0.0803 \\
SN2007on & 29.0764 & 0.0589 \\
SN2008bc & 36.7482 & 0.052 \\
SN2008hv & 30.8722 & 0.0686 \\
SN2008ia & 29.1072 & 0.2377 \\
\\
\hline
\end{tabular}
\end{table}
\newpage
\onecolumn
\begin{longtable}{lccc}\caption{The parameters for the PTF and iPTF SNe in the r-band.}\\
Name & $t_{r_2}$ & $\sigma_{t_{r_2}}^{lower}$ & $\sigma_{t_{r_2}}^{upper}$\\
\hline
\\
PTF10rgn & 0.454 & 0.377 & 0.532 \\
iPTF16fmb & 0.251 & 0.113 & 0.389 \\
iPTF13aol & 0.463 & 0.398 & 0.527 \\
iPTF13dhp & 0.528 & 0.435 & 0.621 \\
iPTF16aas & 0.245 & 0.093 & 0.398 \\
PTF12eac & 0.484 & 0.305 & 0.663 \\
iPTF16gdp & 0.338 & 0.24 & 0.437 \\
iPTF13ani & 0.573 & 0.494 & 0.651 \\
PTF12fxn & 0.36 & 0.176 & 0.545 \\
PTF10hpp & 0.501 & 0.4 & 0.601 \\
PTF10tce & 0.557 & 0.473 & 0.64 \\
PTF10hdn & 0.544 & 0.46 & 0.628 \\
iPTF13ax & 0.505 & 0.448 & 0.561 \\
iPTF16eka & 0.418 & 0.351 & 0.484 \\
PTF10hdm & 0.488 & 0.416 & 0.561 \\
PTF11htb & 0.49 & 0.402 & 0.578 \\
iPTF16hun & 0.293 & 0.089 & 0.498 \\
PTF10nyt & 0.522 & 0.412 & 0.631 \\
PTF10kzf & 0.582 & 0.504 & 0.661 \\
PTF11dec & 0.495 & 0.442 & 0.548 \\
PTF12cjg & 0.447 & 0.259 & 0.635 \\
PTF10urn & 0.459 & 0.355 & 0.563 \\
PTF10uzi & 0.214 & 0.168 & 0.26 \\
PTF12gmf & 0.241 & 0.05 & 0.431 \\
iPTF13ez & 0.49 & 0.438 & 0.541 \\
PTF11blu & 0.524 & 0.459 & 0.589 \\
iPTF14yl & 0.688 & 0.586 & 0.791 \\
PTF10rpt & 0.507 & 0.438 & 0.575 \\
PTF12gmq & 0.444 & 0.246 & 0.642 \\
iPTF14aia & 0.406 & 0.331 & 0.482 \\
PTF11dzm & 0.286 & 0.191 & 0.381 \\
iPTF14apu & 0.341 & 0.182 & 0.5 \\
PTF11kml & 0.478 & 0.367 & 0.588 \\
PTF09alu & 0.315 & 0.223 & 0.407 \\
iPTF16dp & 0.41 & 0.31 & 0.51 \\
PTF10tfs & 0.313 & 0.233 & 0.394 \\
PTF10hcu & 0.435 & 0.331 & 0.539 \\
iPTF13apn & 0.481 & 0.427 & 0.536 \\
PTF11bok & 0.501 & 0.421 & 0.58 \\
PTF10pvi & 0.457 & 0.366 & 0.548 \\
PTF11rnu & 0.356 & 0.229 & 0.483 \\
PTF10hne & 0.493 & 0.426 & 0.56 \\
iPTF16grm & 0.348 & 0.255 & 0.441 \\
iPTF14axv & 0.467 & 0.4 & 0.533 \\
PTF11ivb & 0.395 & 0.32 & 0.47 \\
PTF10hrw & 0.449 & 0.368 & 0.53 \\
PTF11cji & 0.48 & 0.367 & 0.592 \\
iPTF13ckk & 0.527 & 0.471 & 0.583 \\
PTF11bas & 0.482 & 0.42 & 0.543 \\
PTF10ucj & 0.424 & 0.308 & 0.539 \\
PTF12mj & 0.543 & 0.465 & 0.621 \\
PTF10nvh & 0.553 & 0.459 & 0.647 \\
PTF11cmg & 0.365 & 0.271 & 0.46 \\
PTF11cyv & 0.482 & 0.364 & 0.6 \\
PTF11cml & 0.445 & 0.367 & 0.524 \\
iPTF13beg & 0.348 & 0.284 & 0.411 \\
PTF10aayx & 0.394 & 0.284 & 0.503 \\
iPTF14alb & 0.361 & 0.126 & 0.596 \\
PTF10qrj & 0.337 & 0.27 & 0.404 \\
iPTF13dnj & 0.4 & 0.336 & 0.465 \\
iPTF14amb & 0.482 & 0.383 & 0.58 \\
PTF10qnn & 0.471 & 0.379 & 0.564 \\
iPTF14bcl & 0.467 & 0.42 & 0.513 \\
PTF11xe & 0.553 & 0.425 & 0.681 \\
iPTF13adg & 0.478 & 0.4 & 0.556 \\
iPTF13ai & 0.433 & 0.371 & 0.496 \\
PTF12fuu & 0.405 & 0.162 & 0.649 \\
PTF10aajv & 0.302 & 0.265 & 0.339 \\
PTF12kim & 0.519 & 0.393 & 0.644 \\
PTF09dxo & 0.391 & 0.274 & 0.507 \\
iPTF13ccm & 0.482 & 0.443 & 0.521 \\
PTF10mla & 0.458 & 0.337 & 0.578 \\
PTF12dhb & 0.499 & 0.428 & 0.57 \\
PTF10tum & 0.552 & 0.411 & 0.694 \\
iPTF13s & 0.504 & 0.448 & 0.559 \\
PTF12dhk & 0.611 & 0.543 & 0.68 \\
PTF10ucl & 0.413 & 0.301 & 0.526 \\
PTF10qqw & 0.514 & 0.447 & 0.58 \\
PTF12cnl & 0.531 & 0.455 & 0.607 \\
PTF10qkf & 0.47 & 0.397 & 0.542 \\
PTF12keu & 0.546 & 0.416 & 0.676 \\
PTF10glo & 0.578 & 0.544 & 0.612 \\
PTF11qpc & 0.537 & 0.441 & 0.633 \\
PTF10iah & 0.496 & 0.397 & 0.596 \\
PTF10duy & 0.471 & 0.412 & 0.53 \\
iPTF13dkx & 0.525 & 0.471 & 0.579 \\
PTF10rhi & 0.551 & 0.432 & 0.67 \\
PTF10urj & 0.418 & 0.312 & 0.524 \\
PTF12dgy & 0.381 & 0.302 & 0.459 \\
iPTF13czs & 0.47 & 0.417 & 0.523 \\
iPTF13dkl & 0.476 & 0.409 & 0.543 \\
PTF11bof & 0.487 & 0.429 & 0.545 \\
iPTF14afv & 0.441 & 0.367 & 0.515 \\
iPTF13adw & 0.528 & 0.476 & 0.58 \\
iPTF13acz & 0.586 & 0.532 & 0.641 \\
PTF10qsc & 0.549 & 0.46 & 0.638 \\
PTF10egs & 0.42 & 0.33 & 0.51 \\
iPTF13dkj & 0.392 & 0.318 & 0.467 \\
PTF10cmj & 0.497 & 0.351 & 0.644 \\
PTF12cks & 0.555 & 0.507 & 0.603 \\
iPTF14aaf & 0.518 & 0.449 & 0.587 \\
PTF10fxp & 0.398 & 0.308 & 0.488 \\
PTF12juu & 0.372 & 0.221 & 0.523 \\
PTF10hld & 0.497 & 0.428 & 0.567 \\
PTF12hwb & 0.443 & 0.287 & 0.6 \\
iPTF14aik & 0.283 & 0.073 & 0.494 \\
iPTF16gef & 0.297 & 0.145 & 0.449 \\
PTF10hei & 0.446 & 0.374 & 0.519 \\
PTF10qqt & 0.519 & 0.403 & 0.635 \\
PTF10fym & 0.398 & 0.32 & 0.476 \\
PTF10yux & 0.322 & 0.217 & 0.428 \\
PTF10ifj & 0.54 & 0.444 & 0.637 \\
iPTF14axt & 0.361 & 0.282 & 0.44 \\
iPTF13cwq & 0.465 & 0.395 & 0.535 \\
PTF10abws & 0.429 & 0.304 & 0.554 \\
iPTF14yy & 0.376 & 0.314 & 0.437 \\
iPTF13bmn & 0.39 & 0.276 & 0.503 \\
PTF10oum & 0.382 & 0.312 & 0.452 \\
iPTF13cd & 0.429 & 0.363 & 0.495 \\
iPTF16gmw & 0.344 & 0.22 & 0.468 \\
iPTF13dbp & 0.248 & 0.225 & 0.271 \\
PTF12ibh & 0.439 & 0.326 & 0.551 \\
PTF10ivt & 0.388 & 0.271 & 0.505 \\
iPTF13dnh & 0.552 & 0.498 & 0.606 \\
iPTF13cxn & 0.469 & 0.399 & 0.539 \\
PTF11rke & 0.481 & 0.387 & 0.575 \\
PTF12sz & 0.454 & 0.388 & 0.52 \\
iPTF13anh & 0.441 & 0.373 & 0.51 \\
iPTF14aqs & 0.321 & 0.045 & 0.598 \\
iPTF13ag & 0.471 & 0.416 & 0.526 \\
PTF10cko & 0.494 & 0.4 & 0.588 \\
PTF10fxl & 0.556 & 0.513 & 0.599 \\
iPTF13bjb & 0.431 & 0.348 & 0.513 \\
PTF10kiw & 0.526 & 0.407 & 0.645 \\
iPTF13bdb & 0.35 & 0.124 & 0.577 \\
iPTF13dfa & 0.294 & 0.228 & 0.36 \\
PTF10kee & 0.396 & 0.311 & 0.481 \\
PTF10kdg & 0.421 & 0.325 & 0.517 \\
PTF10abkt & 0.32 & 0.17 & 0.47 \\
iPTF14dcd & 0.474 & 0.433 & 0.515 \\
PTF10qkv & 0.469 & 0.419 & 0.519 \\
iPTF13cyy & 0.487 & 0.419 & 0.555 \\
PTF10qly & 0.527 & 0.437 & 0.617 \\
PTF11kaw & 0.491 & 0.399 & 0.583 \\
iPTF13adv & 0.463 & 0.411 & 0.515 \\
iPTF16gua & 0.281 & 0.188 & 0.374 \\
PTF10ujl & 0.414 & 0.259 & 0.569 \\
PTF12gaz & 0.308 & 0.06 & 0.557 \\
PTF11ao & 0.413 & 0.227 & 0.6 \\
PTF10rbp & 0.49 & 0.395 & 0.584 \\
PTF11hfu & 0.588 & 0.451 & 0.725 \\
PTF10feg & 0.561 & 0.483 & 0.64 \\
PTF10goo & 0.498 & 0.44 & 0.555 \\
iPTF13cor & 0.346 & 0.27 & 0.423 \\
iPTF13bun & 0.443 & 0.247 & 0.639 \\
iPTF14anq & 0.408 & 0.203 & 0.614 \\
PTF11rrq & 0.403 & 0.326 & 0.48 \\
iPTF13caz & 0.423 & 0.355 & 0.492 \\
PTF10twd & 0.587 & 0.504 & 0.67 \\
iPTF13cow & 0.477 & 0.39 & 0.563 \\
iPTF13daw & 0.339 & 0.268 & 0.411 \\
PTF10one & 0.37 & 0.302 & 0.438 \\
iPTF16sw & 0.49 & 0.385 & 0.595 \\
iPTF13dni & 0.468 & 0.407 & 0.529 \\
PTF11ilj & 0.537 & 0.406 & 0.667 \\
PTF10qyx & 0.375 & 0.307 & 0.444 \\
iPTF13adm & 0.449 & 0.384 & 0.514 \\
PTF11qvc & 0.516 & 0.419 & 0.614 \\
iPTF16ig & 0.405 & 0.288 & 0.522 \\
PTF10aaju & 0.467 & 0.395 & 0.539 \\
iPTF16fht & 0.208 & 0.073 & 0.344 \\
PTF10xup & 0.473 & 0.411 & 0.534 \\
PTF12kta & 0.402 & 0.306 & 0.499 \\
PTF12csi & 0.493 & 0.423 & 0.563 \\
PTF10lxp & 0.429 & 0.384 & 0.475 \\
PTF10wyq & 0.432 & 0.319 & 0.546 \\
iPTF14fyt & 0.805 & 0.624 & 0.987 \\
PTF10gjx & 0.48 & 0.423 & 0.537 \\
PTF12dxm & 0.388 & 0.146 & 0.63 \\
iPTF13dad & 0.475 & 0.426 & 0.524 \\
PTF12lgq & 0.497 & 0.288 & 0.707 \\
PTF12vr & 0.474 & 0.388 & 0.56 \\
PTF10czc & 0.454 & 0.393 & 0.514 \\
iPTF13ceq & 0.46 & 0.32 & 0.6 \\
PTF10ufj & 0.47 & 0.39 & 0.551 \\
PTF10goq & 0.484 & 0.418 & 0.55 \\
PTF10qwm & 0.539 & 0.447 & 0.631 \\
iPTF13ddg & 0.462 & 0.414 & 0.51 \\
iPTF13akl & 0.487 & 0.388 & 0.586 \\
PTF12gmu & 0.337 & 0.174 & 0.5 \\
PTF10trp & 0.497 & 0.272 & 0.723 \\
PTF11dwn & 0.534 & 0.461 & 0.607 \\
PTF12gqh & 0.443 & 0.357 & 0.529 \\
PTF10abou & 0.513 & 0.425 & 0.601 \\
PTF10gsp & 0.044 & 0.028 & 0.06 \\
iPTF14bjp & 0.514 & 0.46 & 0.567 \\
iPTF13aig & 0.553 & 0.499 & 0.608 \\
PTF12gaw & 0.432 & 0.307 & 0.557 \\
iPTF16fhz & 0.274 & 0.16 & 0.388 \\
\end{longtable}

\bsp	
\label{lastpage}
\end{document}
